
\magnification=\magstep1    
\tolerance=10000
\hsize=15.3 truecm          
\vsize=23 truecm            


\newcount\initpage
\initpage=0
\mathsurround=3pt           
\binoppenalty=10000         
\relpenalty=10000           
\hoffset=0.7 truecm    
\voffset= -1truecm          
\baselineskip=15pt          

\def\cvd{{\hbox{\enspace${\bf \square}$}} \smallskip}
\def\sqr#1#2{{\vcenter{\vbox{\hrule height .#2pt
\hbox{\vrule width .#2pt height#1pt \kern#1pt
\vrule width .#2pt}
\hrule height .#2pt}}}}
\def\square{\ifmmode\mathchoice\sqr54\sqr54\sqr{4.1}3\sqr{3.5}3\else
$\mathchoice\sqr54\sqr54\sqr{4.1}3\sqr{3.5}3$\fi}
\def\1{\vskip .3truecm}
\def\2{\vskip .6truecm}
\def\3{\vskip .9truecm}

\def\die{~}

\def\freccia{\longrightarrow}

\def\ep{\epsilon}
\def\M{{\cal M}}
\def\lor{({\cal M},\langle\cdot,\cdot\rangle)}
\def\connr{D^{\rm R}_s}
\def\freccia{\longrightarrow}
\def\R{I\!\!R}
\def\N{I\!\!N}
\long\def\centro#1{\hfill#1\hfill}
\def\proofend{\ifmmode\eqno\square\else\hfill\square\medbreak\fi}

\def\gr{g_{\scriptscriptstyle{{\rm (R)}}}}

\phantom{a}
\bigskip\bigskip

\centerline{{\bf 
A MORSE THEORY FOR MASSIVE PARTICLES AND PHOTONS
}}
\centerline{{\bf
IN GENERAL RELATIVITY.}}

\1 \bigskip\bigskip

\centerline{{\bf Fabio Giannoni}}
\centerline{Dipartimento di Matematica e Fisica, 
Universit\'a di Camerino}
\centerline{Via Madonna delle Carceri 20--62032--CAMERINO (MC)--ITALY}
\centerline{e-mail: giannoni@campus.unicam.it}

\1 \bigskip\bigskip

\centerline{{\bf Antonio Masiello}}
\centerline{Dipartimento Interuniversitario di Matematica, 
Politecnico di Bari}
\centerline{Via E.Orabona 4 -- 70125--BARI--ITALY}
\centerline{e-mail: masiello@pascal.dm.uniba.it}

\1  \bigskip\bigskip

\centerline{{\bf Paolo Piccione}}
\centerline{Departamento de Matem\'atica, Universidade de S\~ao Paulo}
\centerline{S\~AO PAULO, SP, BRAZIL}
\centerline{e-mail: piccione@ime.usp.br}

\bigskip\bigskip
\bigskip\bigskip

\centro{\vbox{\hsize11truecm\parindent0pt\centerline{\bf Abstract}\medskip
In  this paper we develop a Morse Theory for timelike geodesics 
parameterized by  a constant multiple of proper time. 
The results are obtained  
using an  extension to the timelike case of the relativistic 
Fermat Principle,
and techniques from Global Analysis on infinite dimensional manifolds. 
In the second part of the paper we discuss
a limit process that allows to obtain also a 
Morse theory for light rays. }}

\vfill\break
\noindent
{\bf 1. Introduction}

\1

\noindent
In an arbitrary relativistic space-time, 
modeled by a $4$-dimensional time-oriented Lorentzian
manifold $(\M,g)$, the trajectories of massive objects or massless particles,
like photons,  that move freely under the action of the gravitational
field, are geodesics. 
These geodesics are {\sl timelike\/} in the massive case,
representing the motion of objects traveling slower than the speed of
light, and null, or lightlike, in the case of (massless) 
particles moving at the speed of light. They can be 
characterized by variational principles which can be interpreted as
extensions to General Relativity of the Fermat principle in classical
optics. 

Some of them can be used to describe the so called 
{\sl gravitational lens effect\/} that occurs in Astrophysics whenever 
multiple images of pointlike sources (for example quasars) are
observed (cf. e.g. [SEF]).  In mathematical terminology, a gravitational 
lensing situation can be modeled
in the following way. We consider a Lorentzian manifold 
$(\M,g)$ as a mathematical
model for the spacetime, 
we fix a timelike curve $\gamma$ as the worldline of
a light source and a point $p$ 
as the event where the observation takes place.
Now, the number of images seen by the observer equals 
the number of future pointing lightlike
geodesics from $p$ to $\gamma$. 
Whenever there are two or more such geodesics, we are
in a gravitational lensing situation.  
Alternatively, one could interpret $p$ as an instantaneous pointlike source 
of light and $\gamma$ as the worldline of a receiver.
Since the two problems can be treated in the same way 
from a mathematical point of view,
we shall focus our attention only on this second case. 

It should be remarked that different approaches to the mathematical
modeling of the gravitational lensing effect are possible. For instance, in
 [Pt1,Pt2,Sc],  the authors use a {\sl thin lens\/} approximation; in [L]  
also non thin lenses are considered.  
 
In a recent paper, I. Kovner has suggested a very general version of the
Fermat principle to study timelike and lightlike geodesics (cf. [K]). 
Kovner's principle, justified by plausible arguments
in [K] and rigorously proven by 
V.\ Perlick in [Pe] for the lightlike case, can be stated 
as follows. Among all 
future pointing curves  $z:[0,1]\freccia\M$ 
joining $p$ and $\gamma$ and satisfying $g(z)[\dot z,\dot z]\equiv a$, with
$a\le 0$ fixed, i.e., all possibilities to go from $p$ to $\gamma$ 
at speed less than ($a<0$) or equal to ($a=0$)
the (vacuum) speed of light, 
the geodesics are characterized as stationary points for the
{\sl arrival time\/} (defined using a smooth parameterization of $\gamma$).
In the lightlike case ($a=0$), this principle generalizes
 the Fermat's Principle for light rays in classical optics.

In an absolutely similar fashion,
one could give a time-reversed version of the principle, by interpreting
$p$ as an instantaneous receiver and $\gamma$ the worldline of a source.
In this case, the geodesics are characterized by stationary 
{\sl departure time}.

The aim of this paper is twofold. 
In the first part we shall develop a Morse Theory
for future pointing timelike  geodesics with a  prescribed parameterization
(proportional to the proper time) and joining a
given event with a timelike curve  
in a time--oriented Lorentzian manifold.

In the second part of the article, using a limit process, we shall prove 
the Morse relations for future pointing lightlike geodesic (light rays),
 giving a new and simpler proof with respect to the ones of [GMP1,GMP2], 
where the existence of a smooth time function was assumed.
In this paper we shall only assume the existence of a time--orientation
for the Lorentzian manifold. 
\smallskip

In order to state our results, we now give the basic definitions
and we introduce the notations needed for our setup.  

Let $\lor$ be a time oriented Lorentz manifold and 
let $Y$ be a smooth timelike vector 
field  giving  the time orientation 
(we refer to [BEE,ON] for the basic notions of Lorentzian
Geometry that will be used). We set $m={\rm dim}(\M)$; the
physical interesting case is $m=4$.

Fix an event $p \in \M$ and 
a timelike curve $\gamma\colon\R\freccia\M$. 
On the curve $\gamma$ we shall make the following assumptions: 
\medskip
\centro{\vbox{\hsize13.5truecm%
\item{$\bullet$} $\gamma$ is of class $C^2$;\smallskip
\item{$\bullet$} $\gamma$ is timelike and future pointing;\smallskip
\item{$\bullet$} $\gamma$ is injective;\smallskip
\item{$\bullet$} $\gamma(\R)$ does not contain $p$;\smallskip
\item{$\bullet$} $\gamma(\R)$ is not entirely contained in $I^+(p)$, the 
{\sl causal future\/} of $p$.}\vskip-2truecm\hfill(1.1)\vskip1truecm}
\medskip\noindent
We recall that the causal future of a point $p$ is defined as:
$$\eqalign{I^+(p)=\big\{q\in\M\;\big\vert\;&\ \hbox{\rm there 
exists a future pointing causal curve}\cr
& z:[a,b]\longmapsto\M\ \hbox{\rm\  with $z(a)=p$ and $z(b)=q$}\big\}. }$$
As customary, if $I\subseteq\R$ is any interval,
we will denote by $H^{1,2}(I,\R^n)$
the Sobolev space of all absolutely continuous curves 
$z:I\mapsto\R^n$  having square integrable derivative on $I$.
Given any differentiable manifold $N$, with $n={\rm dim}(N)$,
we define $H^{1,2}([0,1],N)$  as the
set of all  absolutely continuous  curves $z:[0,1]\mapsto N$
such that, for every local chart $(V,\varphi)$ on $N$, with 
$\varphi:U\longmapsto \R^n$
a diffeomorphism, and for every closed subinterval
$I\subseteq[0,1]$ such that $z(I)\subset V$, 
it is $\varphi\circ z\in H^{1,2}(I,\R^n)$. 

It is not difficult to see that this definition of $H^{1,2}([0,1],N)$ may
be given equivalently in the following two ways:
\smallskip
\item{$\bullet$} a curve $z:[0,1]\mapsto N$ belongs to $H^{1,2}([0,1],N)$
if and only if there exists a finite sequence $I_1,\ldots,I_k$ of closed
subintervals of $[0,1]$ and a finite number of charts $\varphi_i:U_i\longmapsto\R^n$
on $N$, $i=1,\ldots,k$, such that $\bigcup_{i=1}^kI_k=[0,1]$, $z(I_i)\subset U_i$, and
$\varphi_i\circ z\in H^{1,2}(I_i,\R^n)$ for all $i=1,\ldots,k$;
\smallskip
\item{$\bullet$} a $C^1$-curve $z:[0,1]\mapsto N$ is in $H^{1,2}([0,1],N)$
if and only if for one (hence for every) Riemannian metric $\gr$ on $N$, the 
integral $\int_0^1\gr({\dot z},\dot z)\;{\rm d}t$ is finite.
\smallskip 
\noindent
A classical result of Global Analysis (see [Pa1]) states that
$H^{1,2}([0,1],N)$ has the
structure of an infinite dimensional manifold, modeled on the Hilbert
space $H^{1,2}([0,1],\R^n)$. Similarly, one defines the Banach manifolds
$H^{k,p}([0,1],N)$, $k\in\N$, $1\le p\le +\infty$, modeled on the Sobolev 
spaces $H^{k,p}([0,1],\R^n)$. In particular, in this paper we will be
concerned with the manifolds $H^{k,p}([0,1],\M)$ and
$H^{k,p}([0,1],T\M)$, where $T\M$ is the tangent bundle of $\M$.

If $\gr$ is any given Riemannian metric on $\M$, for $1\le p\le+\infty$ we also define the
spaces $L^p([0,1],T\M)$ as the set of functions $\zeta:[0,1]\mapsto T\M$
such that the real valued function $\gr(\zeta,\zeta)^{1\over2}$ is in
$L^p([0,1],\R)$. It is easy to see that, by the compactness of $[0,1]$, the
definition of $L^p([0,1],T\M)$ does not depend on the choice of a specific
Riemannian metric $\gr$; observe that $L^p([0,1],T\M)$ does {\sl not\/}
possess any differentiable structure.
\smallskip

The  natural setting to study future pointing 
light rays joining $p$ and $\gamma$ is the following space: 
$$\eqalign{
{\cal L}^+_{p,\gamma}  &=\big\{z:[0,1] \longrightarrow \M \;\big\vert\;
\hbox{ $z \in H^{1,2}([0,1],{\cal M})$, }
\cr
&\qquad\hbox{ 
$\langle Y,\dot z \rangle < 0$ for any 
$s$  such that $\dot z(s)$ exists and it is different from zero, }\cr
&\qquad\hbox{ $\langle \dot z,\dot z \rangle = 0$ a.e.,  
$z(0) = p$, $z(1) \in\gamma(\R)$ }\big\}.\cr}
$$ 
Here the $H^{1,2}$--regularity is used 
because it is the simplest one if we want  to give an infinite dimensional 
approach to the Morse Theory. 

Unfortunately,
${\cal L}^+_{p,\gamma}$ is not a $C^1$-submanifold of $H^{1,2}([0,1],\M)$,
but it only has a Lipschitz regularity. 
For this reason we shall approximate it by  the family of smooth submanifolds 
of $H^{1,2}([0,1],\M)$, parameterized by a positive number
$\epsilon$, given by
$$\eqalign{
{\cal L}^+_{p,\gamma,\ep} =&  
\big\{z:[0,1] \mapsto \M\;\big\vert\; 
 z \in H^{1,2}([0,1],{\cal M}),\   
\langle Y(z),\dot z\rangle < 0\ {\rm a.e.},  \cr
&\quad\hbox{ 
$\langle \dot z,\dot z \rangle = - \ep^2$ a.e.,  
$z(0) = p$, $z(1) \in \gamma(\R)$ 
 } \big\}.\cr}
$$ 
To complete our variational framework  we introduce the  
{\sl arrival time\/} functional $\tau$ 
which  assigns to each curve ending on $\gamma$
the value of the parameter of $\gamma$ at the arrival point.
The functional $\tau$ is defined on the manifold:
$$\displaylines{
\Omega^{1,2}_{p,\gamma} = 
\big\{ z:[0,1] \longrightarrow \M\;\big\vert\; z  \in  H^{1,2}, \cr
\qquad \quad z(0)=p,\ z(1) \in \gamma ({I\!\!R})\big\},\cr}
$$ 
as
$$
\tau(z) = \gamma^{-1}(z(1)).
$$
Observe that $\tau$ is well defined because $\gamma$ is injective.
 
Some relativistic versions of the Fermat Principle 
have  been already used 
(cf.\ e.g.\ [GMP2] and the reference therein) to develop a Morse Theory
for light rays. 
However, the Morse Relations for timelike geodesics with 
prescribed parameterization
has not been obtained yet. 
Moreover, the results for light rays 
in [GMP1,GMP2] 
have been proven under the extra assumption of stable causality
for $\M$, i.e.,  assuming the existence
of a smooth global time function $T:\M\longmapsto\R$ on $\M$,
and using the following functional 
$$
Q(z) = \int_0^1 
\langle \dot z ,\nabla T \rangle^2\die ds,
$$
where $\nabla T $ is the Lorentzian gradient of $T$. 

In spite of the analogy with the energy functional in Riemannian
manifolds, the  critical  points  of $Q$ 
on the approximating manifolds ${\cal L}^{+}_{p,\gamma,\ep}$
do not have a clear geometrical or physical meaning; moreover,  the  
Euler-Lagrange equations for the Lagrangian function of $Q$
are very complicated. 
This is  one of the main reasons making the proof of Morse theory 
in [GMP2] quite involved.

In  this  paper, thanks to the use of the  arrival time functional $\tau$ 
on the manifolds ${\cal L}^{+}_{p,\gamma,\ep}$,
we  first obtain the Morse Relations for the timelike geodesics, then, 
using a limit process 
as $\ep \to 0$,  we extend the results to
the case of lightlike geodesics. 
 
In order to avoid technical difficulties that could make not 
completely clear the  advantages  of  this  new  approach,  
we will consider only the case where $\M$ is a manifold without 
boundary. It is worthy to observe here that the techniques
presented in this paper can be employed also in the study
of causal geodesics in manifolds having a causally convex boundary.

Before stating the main results of the present paper, let us  
recall the notions of conjugate point along a geodesic and 
the notion of  geometric index.

We denote by $D$ the Levi--Civita connection of the metric $g$;
moreover, let $R$ be the {\sl curvature tensor\/} of $g$, defined
with the following sign convention:
$$R(X,Y)Z=D_XD_YZ-D_YD_XZ-D_{[X,Y]}Z,$$
where $X, Y, Z$ are vector fields on $\M$.
\2

\noindent
{\bf Definition  1.1.}\die 
Let $z\colon [0,1]\longrightarrow \M$ be a geodesic.
The point $z(s)$, $s \in ]0,1]$ is said to be {\it conjugate} to $z(0)$ 
along $z$ if there exists a non zero smooth vector field $\zeta$ along 
$z_{|[0,s]}$ (called {\it Jacobi field}), such that
$$
D^2_s\zeta + R(\zeta,\dot z)\dot z = 0~,
\eqno(1.2)
$$
and satisfying the boundary condition
$$
\zeta(0) = 0, \quad \zeta(s) = 0\die .
\eqno(1.3)
$$
The {\it multiplicity} of the conjugate point $z(s)$ is the maximal number
of linearly independent Jacobi fields satisfying (1.3).
The {\it geometric index} $\mu (z)$ of the geodesic $z$ is the number of 
points $z(s)$ conjugate to $z(0)$ along $z$, counted with their multiplicity.

\2

We  shall prove that the functional $\tau$ on 
${\cal L}^+_{p,\gamma,\epsilon}$ is of class 
 $C^2$. 
Let $z$ be a critical point of $\tau$ on ${\cal L}^+_{p,\gamma,\epsilon}$.
The Morse index $m(z,\tau)$ is defined 
as the maximal dimension of a
subspace of $T_z{\cal L}^+_{p,\gamma,\epsilon}$
(the tangent space to ${\cal L}^+_{p,\gamma,\epsilon}$ at $z$), where the 
Hessian of $\tau$ at $z$ is negative definite.

The first result concerns the Fermat principle in 
${\cal L}^+_{p,\gamma,\ep}$.

\2

\noindent
{\bf Theorem  1.2.}\die 
{\it A curve 
$z$  is a critical point of $\tau$ on ${\cal L}^+_{p,\gamma,\ep}$ 
if and  only if $z$ is a future pointing timelike geodesic 
joining $p$ with $\gamma$ such that $\langle\dot z,\dot z\rangle = -\ep^2$.
Moreover, if $z(1)$ is nonconjugate to $p = z(0)$ along $z$, then
$m(z,\tau) = \mu(z)$.}

\2

To write Morse Relations for $\tau$ in 
${\cal L}^+_{p,\gamma,\ep}$,  
(where $\ep > 0$ is fixed), 
we need to assume that $\lor$ is {\sl strongly causal}. 
This means that, for any point $q\in\M$, there is no future pointing
causal curves  starting arbitrarily close to $q$, leaving some fixed
neighborhood of $p$ and returning arbitrarily close to $q$ (cf.\ [BEE,ON]).

Moreover we need to recall some topological definitions. 
Let $X$ be a topological space, ${\cal K}$ an algebraic field, for any $q
\in \N$ we denote by
$H_q(X,{\cal K})$ the $q$--th homology group of $X$with coefficient in
${\cal K}$.
Since ${\cal K}$ is a field, $H_q(X,{\cal K})$ is a vector space. 
The dimension $\beta_q(X,{\cal K})$ of $H_q(X,{\cal K})$ is called
$q$--the Betti number of $X$ (with coefficients in ${\cal K}$. 
Finally the Poincar\'e polynomial of $X$ is the formal series with
coefficients in $\N\cup+\infty$ defined as
$$
{\cal P}_r(X,{\cal K}) = 
\sum_{q = 0}^\infty
\beta_q(X,{\cal K}) .
$$

\2

\noindent
{\bf Theorem 1.3.}\die 
{\it  Let $\lor$ be strongly causal, $\gamma$ a curve in
$\M$ satisfying (1.1) and:}

\smallskip
\item{1)} ${\cal L}^+_{p,\gamma,\ep} \not= \emptyset$;

\item{2)}
{\it for  any  geodesic  
$z  \in {\cal L}^+_{p,\gamma,\epsilon}$, $z(1)$ is nonconjugate to 
 $z(0) = p$.}

\item{3)} {\it 
the functional $\tau$ is pseudo--coercive on 
${\cal L}^+_{p,\gamma,\ep}$, namely:
for  any  $c \in {I\!\!R}$,  
there exists $K_{c}$ compact subset of $\cal M$,
such that $z([0,1]) \subset K_{c}$ for any 
$z \in {\cal L}^+_{p,\gamma,\ep}$ satisfying
$\tau(z) \leq c$}.
\smallskip

\noindent
{\it Then, for any coefficient field ${\cal K}$, there exists a formal
series $S(r)$ with coefficients in ${I\!\!N}\cup \{+\infty\}$, such that}
$$
\sum_{z\in{\cal G}_{p,\gamma,\ep}^+} r^{\mu(z)} = 
{\cal P}_r({\cal L}_{p,\gamma,\ep}^+;{\cal K}) + (1+r)S(r).
\eqno(1.4)
$$

\2

\noindent
Here ${\cal G}_{p,\gamma,\ep}^+$ is the set of the timelike geodesics in 
${\cal L}_{p,\gamma,\ep}^+$.

\2

\noindent
{\bf Remark 1.4.}~
The set of assumptions (1.1) on the curve $\gamma$ imply immediately
that  $\tau$ is bounded from below in ${\cal L}_{p,\gamma,\ep}^+$.

\2

\noindent  
Taking  the  limit  as  $\ep \to 0$,  we  obtain also the Morse 
 relations for light rays. 

\2

\noindent
{\bf Theorem 1.5.}\die 
{\it Let $\lor$ be strongly causal, $\gamma$ a curve satisfying (1.1) and:}

\item{1)} 
${\cal L}^+_{p,\gamma} \not=\emptyset;$

\item{2)} 
{\it for any geodesic 
$z \in {\cal L}^+_{p,\gamma}$,  $z(1)$ is nonconjugate to $z(0) = p$.}

\item{3)}
{\it 
the functional $\tau$ is pseudo--coercive on 
${\cal L}^+_{p,\gamma}$, namely:
for  any  $c \in {I\!\!R}$,  
there exists $K_{c}$ compact subset of $\cal M$,
such that $z([0,1]) \subset K_{c}$ for any 
$z \in {\cal L}^+_{p,\gamma}$ satisfying
$\tau(z) \leq c$}.
\smallskip

{\it Then, for any coefficient field ${\cal K}$, there exists a formal
series $S(r)$ with coefficients in ${\N}\cup \{+\infty\}$, such that}
$$
\sum_{z\in{\cal G}_{p,\gamma}^+} r^{\mu(z)} = 
{\cal P}_r({\cal L}_{p,\gamma}^+;{\cal K}) = (1+r)S(r).
\eqno(1.5)
$$

\2

\noindent
Here ${\cal G}_{p,\gamma}^+$ is the set of the lightlike geodesics in 
${\cal L}_{p,\gamma}^+$.

\2

\noindent
For the limit process the following results are crucial.

\2

\noindent
{\bf Theorem 1.6.}\die
{\it Assume that $\lor$ is strongly causal and 
$\tau$ is pseudo--coercive on ${\cal L}^+_{p,\gamma}$. 
Let  $c \in {\R}$, $(\ep_m)_{m\in{\N}}$ 
any sequence in $\R^+$ with $\ep_m \to 0$,
and   $(z_m)_{m\in {\N}}$ a sequence
of (timelike) geodesics in  
${\cal L}^+_{p,\gamma,\ep_m}$, satisfying 
$\tau(z_m) \leq c$ for all $m\in{\N}$. 
Then,  
$z_m$ has a subsequence which is convergent (with respect to 
the $C^2$-norm) to a future pointing lightlike geodesic joining $p$ and
$\gamma$.}

\2

\noindent
{\bf Theorem 1.7.}\die  
{\it 
Let $(z_m)_{m\in{\N}}$  be a sequence of timelike geodesics
convergent with respect to the $C^2$--norm to a lightlike geodesic $z$, 
such that $z(0)$ and $z(1)$ 
 are non conjugate. Then:}
$$
\mu(z_m) = \mu(z) \quad\quad 
\hbox{ for any $m$ sufficiently large }~.
$$

\2

\noindent
Theorem 1.6 will be proved in section 6. The proof of Theorem 1.7 
involves the notion of {\sl Maslov\/} index for a semi-Riemannian
geodesic (see~[H, MPT]).  For causal Lorentzian geodesics, the Maslov
index coincides with the geometric index of the geodesic, while 
in the general case it is given by a sort {\sl algebraic\/} count of the
multiplicities of the conjugate points along the geodesic. The Maslov index
can be characterized as the intersection number between a curve and a
codimension one subvariety of the {\sl Lagrangian Grassmannian\/} of a symplectic
space, and thus it is stable by homotopies. The stability of the geometric index
can be proven in more general contexts; details of the proof may be found in 
[MPT].
\1
The Morse  Relations  provide a global 
description of the multiple image
effect   for  pointlike  sources.  
Some  information  about  the  physical  
phenomenon  can be obtained directly using them: 
for instance the information about the  odd  number  of images 
predicted  by astrophysicists (cf. [GMP2, Theorem 1.16]). 

\vfil
\eject

%
%
%
%
%
\noindent
{\bf 2. Existence of minimizers}

\1

\noindent
Fix $\ep > 0$.
In order to develop a Morse Theory on ${\cal L}_{p,\gamma,\ep}^+$ 
using the functional $\tau$, we should need the Palais--Smale condition
for 
$\tau$ on ${\cal L}_{p,\gamma,\ep}^+$.
Namely, we should need that any sequence $(z_m)_{m\in\N}$ such that 
$\tau(z_m)_{m\in\N}$ is uniformly bounded with respect to $m$ and 
${\rm d}\tau (z_m)\to 0$ as $m \to \infty$, had a converging subsequence
in ${\cal L}_{p,\gamma,\ep}^+$.
Unfortunately, $\tau$ has homogeneity 1 as a length functional on a
Riemannian manifold (as it can be proved using local coordinates). 
Therefore the natural space to study the Palais--Smale condition is the
space 
$$\eqalign{
\hat{\cal L}_{p,\gamma,\ep}^+ = 
\big\{z \in H^{1,1}(&[0,1],\M)\;\big\vert\; 
\langle\dot z,\dot z\rangle = -\ep^2 \hbox{
a.e. }, \cr
 &\langle Y(z),\dot z\rangle < 0 \hbox{ a.e. }, 
z(0) = p, z(1) \in \gamma({I\!\!R})\big\}~,}\eqno{(2.1)}
$$
where $H^{1,1}([0,1],\M)$ denotes the space of the absolutely continuous
curves (on any local chart) whose first derivative is integrable.

But to develop a Morse Theory it is really more convenient to work on the
Hilbert manifold ${\cal L}_{p,\gamma,\ep}^+$. 
For this reason we shall use a curve shortening procedure, working on the
curve space ${\cal L}_{p,\gamma,\ep}^+$.
The space ${\cal L}_{p,\gamma,\ep}^+$ is equipped with a structure of 
 infinite dimensional
manifold and its tangent space at a point $z$ is given by
$$\displaylines{
T_z{\cal L}_{p,\gamma,\ep}^+ = 
\{\zeta \in H^{1,2}([0,1],T{\cal M}):\zeta(0) = 0, 
\zeta(1) \parallel \dot\gamma(z(1))\die ,\cr
\hfill
\langle z,D_s\zeta\rangle = 0 \hbox{ a.e.}, 
\zeta(s) \in T_{z(s)}\M \hbox{ for any $s \in [0,1]$ } \}\die , 
\hfill\llap{(2.2)}}
$$
where $T\M$ is the tangent bundle of $\M$ (cf. [GMP1] 
replacing there $\nabla T$ by $Y$).

We introduce a Riemannian structure on $\M$ setting for any $p \in \M$
and $\zeta \in T_p\M$, 
$$
\langle \zeta,\zeta\rangle_{\rm (R)} = 
\langle \zeta,\zeta\rangle - 
2{{\langle \zeta,Y(z)\rangle^2}
\over {\langle Y(z),Y(z)\rangle}}\die.
\eqno(2.3)
$$
The {\sl wrong way Schwartz's inequality\/} (cf.\ [ON]) shows that (2.3) is a Riemannian
structure on $\M$. 
We shall denote by $d_{\rm R}$ the distance function induced by (2.3).

A Riemannian structure can be introduced 
on the manifold ${\cal L}_{p,\gamma,\ep}^+$, setting for any 
$z \in {\cal L}_{p,\gamma,\ep}^+$ and 
$\zeta \in T_z{\cal L}_{p,\gamma,\ep}^+$, 
$$
\langle\zeta,\zeta\rangle_1 = 
\int_0^1\langle D_s\zeta,D_s\zeta\rangle_{\rm (R)}\die {\rm d}s\die,
\eqno(2.4)
$$
The proof is formally the same as in [GMP1], where the existence of a time
function is assumed.

Now, for any $[a,b] \subset [0,1]$, 
$-\infty < \alpha < \beta < +\infty$, 
$q\in\M$ and $\delta\colon ]\alpha,\beta[ \freccia \M$ smooth timelike
curve, we set
$$\displaylines{
{\cal L}_{q,\delta,\ep}^+([a,b])
 = \{z \in H^{1,2}([a,b],\M:
 z(a) = q, z(b) \in \delta(]\alpha,\beta[)\die ,\cr
\hfill
\langle\dot z,\dot z\rangle = -\ep^2 \hbox{ a.e.},
\langle\dot z,Y(z)\rangle < 0 \hbox{ a.e. }\}\die .
\hfill\llap{(2.5)}
}
$$
Note that  $\delta$ is injective because $\M$ is strongly causal.

The main result of this section is 
the following result on the existence and the uniqueness
of minimizers of the
arrival time $\tau$ between a point and a "sufficiently close" integral
curve $\delta$ of the vector field $Y$ (obviously $\delta$ is a timelike
curve). 

\2

\noindent
{\bf Theorem 2.1.}\die
{\it Fix $-\infty < \alpha < \beta < +\infty$. For any $q \in \M$ there
exists a positive number $\rho(q)$ having the following property:}

{\it For any integral curve $\delta:]\alpha,\beta[ \freccia \M$ of $Y$
such that $d_{\rm R}(q, \delta({{\alpha + \beta} \over 2})) \leq \rho(q)$, and for any
interval $[a,b]$ such that $0< |b-a|\leq \rho(q)$, 
there exists one and only one  
$z \in {\cal L}_{q,\delta,\ep}^+([a,b])$ which minimizes the arrival time
on ${\cal L}_{q,\delta,\ep}^+([a,b])$}

\2

\noindent
Note that in the statement of theorem 2.1, 
the arrival time is given by $\tau(z) = \delta^{-1}(z(b))$. Set 
$$\displaylines{
T_z\hat {\cal L}_{q,\delta,\ep}^+([a,b]) = 
\{\zeta \in H^{1,1}([a,b],T{\cal M}: \zeta(a) =0, \zeta(b)\parallel 
\dot\delta(z(b))\cr
\hfill
\langle z,D_s\zeta\rangle = 0 \hbox{ a.e.}, 
\zeta(s) \in T_{z(s)}\M, \hbox{ for any $s \in [0,1]$ } \}\die. 
\hfill\llap{(2.6)}\cr}
$$
Note that the space $T_z\hat {\cal L}_{q,\delta,\ep}^+([a,b])$ must be
considered as a tangent space, but only in a "Gateaux" sense. 
This is what we need to prove Theorem 2.1.

In order to prove Theorem 2.1, some preliminary results are needed. 
The first says that $\tau$ satisfies the Palais--Smale condition with
respect to the {\it admissible variations} in 
$T_z\hat {\cal L}_{q,\delta,\ep}^+([a,b])$ and with respect to the
"Finsler" 
structure on $\hat {\cal L}_{q,\delta,\ep}^+([a,b])$ defined in the
following way: for any $z \in \hat {\cal L}_{q,\delta,\ep}^+([a,b])$ and
for any $\zeta \in T_z\hat {\cal L}_{q,\delta,\ep}^+([a,b])$, we set
$$
\|\zeta\|_{1,a,b}\equiv \|\zeta\| = 
\int_a^b
\left(\langle D_s^{\rm R}\zeta,D_s^{\rm R}\zeta\rangle_{\rm (R)} 
+\langle \zeta,\zeta\rangle_{\rm (R)}\right)^{1/2}\die{\rm d}s\die ,
\eqno(2.7)
$$
where $D_s^{\rm R}$ denotes the Levi--Civita connection with respect to
the Riemannian metric (2.3). 

\2

\noindent
{\bf Remark 2.2.}\die
Note that since $\delta$ is a curve of class $C^2$ and $\tau$ is
characterized by the relation $\delta(\tau(z)) = z(b)$, we have that $\tau$
is a functional of class $C^2$ on the space of the curves parameterized on
the interval $[a,b]$ and joining $q$ and $\delta$. 
Moreover its differential along a direction $\zeta$ is given by
$$
\dot\delta(\tau(z)){\rm d}\tau(z)[\zeta] = \zeta(b)\die.
$$
Therefore
$$
{\rm d}\tau(z)[\zeta] = 
{
{\langle\dot\delta(\tau(z)),\zeta(b)\rangle}
\over
{\langle\dot\delta(\tau(z)),\dot\delta(\tau(z))\rangle}
}
\die.
\eqno(2.8)
$$

\2

\noindent
{\bf Remark 2.3.}\die
In the rest of the paper it will be often used the parallel transport of 
$\dot\delta(z(b))$ along $z$, namely the solution $U(z)$ of the Cauchy
problem 
$$\cases{
D_s U(z) = 0\cr
U(b) = \dot\delta(\tau(z))\cr}
\eqno(2.9)
$$
where $D_s$ is the covariant derivative along $z(s)$. 
Note that if $z$ has a $H^{1,r}$--regularity, then also $U(z)$ is of class
$H^{1,r} (r \in [1,\infty])$.

\2

\noindent
Since the parallel transport is an isometry, the vector field $U(z)$ along
$z$ is timelike and for any $s\in [a,b]$, 
$$
\langle\dot\delta(\tau(z)),\dot\delta(\tau(z))\rangle  = 
\langle U(z)(s),U(z)(s)\rangle\die .
$$
Moreover, any vector field $\zeta$ along $z$ such that $\zeta(a) = 0$, 
$\zeta(b) = 0$ can be projected on
$T_z\hat {\cal L}_{q,\delta,\ep}^+([a,b])$ using $U(z)$. Indeed, set
$$
V_\zeta (s) = \zeta(s) - \mu(s)U(z)(s), \qquad\qquad
\mu(s) = \int_a^s
{
{\langle D_s\zeta,\dot z\rangle}
\over
{\langle U(z),\dot z\rangle}
}\die{\rm d}r\die .
\eqno(2.10)
$$
Clearly $V_\zeta \in 
T_z\hat {\cal L}_{q,\delta,\ep}^+([a,b])$
and, by (2.8), 
$$
{\rm d}\tau(z)[V_\zeta] = 
{\rm d}\tau(z)[\zeta - \mu U(z)] = 
-\mu(b) = -
\int_a^b
{
{\langle D_s\zeta,\dot z\rangle}
\over
{\langle U(z),\dot z\rangle}
}\die{\rm d}r\die .
\eqno(2.11)
$$
Note that $0$--homogeneity of the map
$$
\theta \freccia {\theta\over  {\langle U(z),\theta\rangle}}
$$
shows that the vector field 
$\langle U(z),\dot z\rangle^{-1}\dot z$ is uniformly bounded, and
therefore $\mu(s) \in H^{1,1}([a,b],\R)$.

\2

\noindent
{\bf Proposition 2.4.}\die
{\it Let $(z_m)_{m \in \N}$ be a sequence of curves of class $C^1$ and
such that $z_m \in \hat {\cal L}_{q,\delta,\ep}^+([a,b])$ for any
$m\in\N$. 
Assume that:}


\item{(i)} 
{\it $\tau(z_m) \to c \in ]\alpha,\beta[$, as $m \to \infty$, where
 $]\alpha,\beta[$ is the interval where $\delta$ is defined;
}

\item{(ii)}
{\it 
$\sup
\{
|{\rm d}\tau(z_m)[\zeta]|:\zeta \in 
T_{z_m}\hat {\cal L}_{q,\delta,\ep}^+([a,b]), 
\|\zeta\|_{a,b,1} \leq 1\} \to 0$, as $m\to \infty$.}

{\it Then the sequence $(z_m)_{m \in \N}$ contains a subsequence
converging to a curve $z$ with respect to the $C^1$--norm}.

\2

\noindent
In order to prove Proposition 2.4, the following remarks and lemmas are
needed. 

\2

\noindent
{\bf Remark 2.5.}~ 
It is not difficult to verify that for any $z_0 \in \M$
there exists a local chart $(U,\varphi)$ of $\M$ containing $z_0$ such that  
$\varphi(U) = V \times I$, where $V$ is a convex open subset of
$\R^{n},~n=m-1$, $I$ is an open interval,
$$\displaylines{
\varphi(U) = \{(x,t):
x = (x_1,\dots,x_n), \cr
\hbox{{\it the distribution generated by the}} \quad
{{\partial}\over{\partial x_i}}\hbox{{\it 's is spacelike and}} \quad
{{\partial}\over{\partial t}} = Y\},\cr}
$$
and the Lorentzian metric $g$ on $\varphi(U)$ can be written as 
$$
{ds}^{2} =
{\langle\alpha(x,t)\xi,\xi\rangle}_0 +
2{\langle\Gamma(x,t),\xi\rangle_0\theta} -
\beta(x,t)\theta^2~
\eqno(2.12)
$$
where $\langle {\cdot},{\cdot}\rangle_{0}$
is a Riemann structure on $V$, $\alpha(x,t)$ is a positive linear operator,
$\Gamma$ is a 
smooth vector field, $\beta(x,t)$ is a smooth positive scalar field,
and $(\xi,\theta) \in {\R}^n \times \R.$

\2

\noindent
{\bf Lemma 2.6.}
{\it Assume that $\tau$ is pseudocoercive on 
$\hat{\cal L}^+_{p,\gamma,\ep}$ (or equivalently on 
${\cal L}^+_{p,\gamma,\ep}$). 
Then, for any $c \in {\bf R}$ there exists $D(c) > 0$ such that}
$$
\tau(z) \leq c \Longrightarrow 
\int_0^1 \sqrt{\langle\dot z,\dot z\rangle_{\rm R}}{\rm d}s \leq D(c)~.
$$

\1

\noindent
{\bf Proof.}~
Assume by contradiction that there exists a sequence 
$(z_m)_{m\in{\N}}$ in $\hat{\cal L}^+_{p,\gamma,\ep}$ such that 
$\tau(z_m)\leq c$, for any $m \in {\N}$ and
$$
\int_0^1 \sqrt{\langle\dot z_m,\dot z_m\rangle_{\rm R}}{\rm d}s 
\to +\infty~.
\eqno(2.13)
$$
Set $\hat z_{m}(s) = z_{m}({s \over \lambda_{m}})$ where
$\lambda_m = \sup \{\langle \dot z_{m}, \dot z_{m} \rangle_{R}^{1/2}: s \in [0,1] \}$.  
By pseudocoercivity (and Ascoli-Arzel\'a's Theorem), up to passing to a subsequence 
there exists a
curve $z\colon \R^{+} \freccia \M$ such that
$$
\hat z_m \freccia z \quad \hbox{ uniformly on the compact subsets of $\R^{+}$ }~,
\eqno(2.14)
$$
 
Now fix $r > 0$ and consider the interval $[0,r]$. Suppose that $z(r+1)$ does not intersect $z([0,r])$. (Since $\M$ is strongly causal and any $\hat z_m $ is causal, this means that 
$z$ is not constant on the interval $[r,r+1]$). The strongly causality of $\M$ implies (arguing by contradiction) that $z(s)$ is uniformly far from $z([0,r])$ on $[r+1,+\infty[$. Therefore we can use a countable set of local charts  $(U_j,\varphi_j), j = 1,...,k$ as in Remark 2.5 and 
the $t$--coordinate on any $\varphi_j(U_j)$ to construct, 
without ambiguity,  a smooth map $T$ on
a relatively compact neighborhood 
${\cal U}$ of $z(\R^{+})$, such that for any $q \in {\cal U}$, 
$$
\langle\nabla T(q),\nabla T(q)\rangle < 0  \quad\hbox{and} \quad 
\langle\nabla T(q),Y(q)\rangle < 0~.
$$
Now any $z_m$ is timelike and $\langle Y(z_m),\dot z_m\rangle < 0$ 
for $m$ and for  any $s \in [0,1]$. Then, for any $m$ sufficiently large, 
$\langle\nabla T(z_m),\dot z_m\rangle > 0$ for any $s \in [0,1]$.
Moreover since $\tau(z_m) \leq c$, we have 
(unless to consider a subsequence) that
$$
T(z_m(1)) \hbox{ is bounded }~.
$$
Now, 
$$
T(z_m(1)) - T(p) = 
T(z_m(1)) - T(0) = 
\int_0^1\langle \nabla T(z_m),\dot z_m\rangle{\rm d}s~,
\eqno(2.15)
$$
while, by (2.3) and 
the choice of the orientation of $\nabla T (z)$,
there exists $\nu_0$ such that 
$$
\langle\nabla T(z_m),\dot z_m\rangle \geq
\nu_0
\sqrt{\langle \dot z_m,\dot z_m\rangle_{\rm (R)}}
\eqno(2.16)
$$
for any $s \in [0,1]$ and $m$ sufficiently large (recall that
$z_m \in \hat{\cal L}^+_{p,\gamma,\ep}).$

Since $T(z_m(1))$ is bounded, combining (2.14)--(2.15)
gives the boundedness of 
$$
\int_0^1\sqrt{\langle\dot z_m,\dot z_m\rangle_{\rm (R)}}~ds~,
$$
in contradiction with (2.13).\cvd

\2

\noindent
{\bf Proof of Proposition 2.4.}~
The proof will be carried out assuming $[a,b] = [0,1]$.
Since $(z_m)_{m\in{\N}}$ is a Palais--Smale sequence, 
$$
\lim_{m\to\infty}
\left(
\sup\big\{\|\tau'(z_m)[\zeta]\|_1:
\zeta\in T_{z_m} \hat{\cal L}^+_{p,\gamma,\ep},\|\zeta\|_1\leq 1\big\}
\right) = 0 ~.
$$
By assumptions (i) and pseudocoercivity, there exists $K$, compact subset
of $\M$ such that 
$$
z_{m}([0,1]) \subset K  \hbox{ for any m. }
$$
Moreover well known results on dual Sobolev spaces (cf. [Br]) imply that
$$
\tau'(z_m)[\zeta] = 
\int_0^1 \langle\alpha_m,\connr \zeta\rangle_{\rm (R)}~{\rm d}s + 
\int_0^1 \langle\beta_m,\zeta\rangle_{\rm (R)}~{\rm d}s~,
\eqno(2.17)
$$
where $\alpha_m$ and $\beta_m$ are 
$L^{\infty}$--vector fields along $z_m$ and 
$$
\alpha_m \freccia 0,
\quad\quad
\beta_m \freccia 0
\quad \hbox{ uniformly }~.
$$
Now,
$$
\connr\zeta - D_s\zeta = \Gamma(z_m)[\dot z,\zeta]~,
$$
where $\Gamma(z_m)$ is a bilinear map, whose components are smooth
functions of $z_m$.
Then there exists 
a vector field $\hat\beta_m$ along $z_m$ (of class $H^{1,1}$)
and a bilinear map $B(z_m)[\cdot,\cdot]$ such that,
$\hat\beta_m \to 0$ uniformly and
$$
\tau'(z_m)[\zeta] = 
\int_0^1 \langle\alpha_m,D_s\zeta\rangle~{\rm d}s + 
\int_0^1 \langle B(z_m)[\alpha_m,\dot z_m],\zeta\rangle~{\rm d}s +
\int_0^1 \langle\hat\beta_m,\zeta\rangle~{\rm d}s~.
\eqno(2.18)
$$
Then, if $\mu$ and $U(z_m)$ are as in (2.9)--(2.10), 
with $\zeta$ replaced by 
$W$, for every $W \in C^\infty_0([0,1],T\M)$ such that 
$W(s) \in T_{z_m(s)}\M$ for any $s$, we have:
$$\displaylines{
\tau'(z_m)[W-\mu U(z_m)] = \cr
\int_0^1 \langle\alpha_m,D_s(W-\mu U(z_m))\rangle~{\rm d}s + 
\int_0^1 
\langle B(z_m)[\alpha_m,\dot z_m] + 
\hat\beta_m,W-\mu U(z_m)\rangle{\rm d}s~.\cr}
$$
Since $D_s U(z_m) = 0$ and 
$$
\mu(s) = 
\int_0^s
{
{\langle D_s W,\dot z_m\rangle}
\over
{\langle U(z_m),\dot z_m\rangle}
}
~{\rm d}s~,
$$
by  (2.10) we have:
$$\displaylines{
-\int_0^1
\langle D_s W,
{
{\dot z_m}
\over
{\langle U(z_m),\dot z_m\rangle}
}
\rangle~{\rm d}s = \cr
\int_0^1 
\langle D_s W - 
{
{\langle D_s W,\dot z_m\rangle}
\over
{\langle U(z_m),\dot z_m\rangle}
}
U(z_m),
\alpha_m\rangle~{\rm d}s
+
\int_0^1
\langle B(z_m)[\alpha_m,\dot z_m] + \hat\beta_m,W\rangle~{\rm d}s 
\cr
-
\int_0^1 \!\!
\int_0^s
\left(
{
{\langle D_\sigma W,\dot z_m\rangle}
\over
{\langle U(z_m),\dot z_m\rangle}
}
~{\rm d}\sigma~
\right)
\langle B(z_m)[\alpha_m,\dot z_m]\hat\beta_m,U(z_m)\rangle~{\rm d}s~.
\cr}
$$
Now,  since $z_m([0,1]) \subset K$ for all $m \in {\N}$, 
$$ 
U(z_m) \quad
\hbox{ and }
{
{\dot z_m}
\over
{\langle U(z_m),\dot z_m\rangle}
}
\quad
\hbox{ are uniformly bounded }~.
$$
Moreover, by Lemma 2.6 the sequence $(\dot z_m)_{m\in{\N}}$ is bounded 
in $L^1([0,1],T\M)$. Since $\alpha_m$ and $\hat \beta_m \to 0$ uniformly, 
the covariant primitive 
$$
\int_0^s 
\left(
B(z_m)[\alpha_m,\dot z_m] + \hat \beta_m
\right)~{\rm d}\sigma
$$
tends uniformly to 0. 
Therefore, an integration by parts shows the existence of a vector field
$A_m$ along $z_m$, such that $A_m$ tends uniformly to 0 and 
$$
\int_0^1 
\langle D_s W,
{
{\dot z_m}
\over
{\langle U(z_m),\dot z_m\rangle}
}
\rangle~{\rm d}s +
\int_0^1
\langle D_s W,A_m\rangle~{\rm d}s = 0~,
$$
for any vector field 
$W \in C^\infty_0([0,1],T\M)$ such that 
$W(s) \in T_{z_m(s)}\M$ for any $s$.

The arbitrariness of $W$ gives the existence of a vector field 
$Z_m \in T_{z_m} \hat{\cal L}^+_{q,\delta,\ep}$ such that
$$
D_s Z_m = 0 \quad\quad 
\hbox{ and }
{
{\dot z_m}
\over
{\langle U(z_m),\dot z_m\rangle}
}
+ A_m = Z_m~.
\eqno(2.19)
$$
Since $D_s Z_m = 0$, the function $C_m = \langle Z_m,Z_m\rangle$ is
constant. Moreover, since $\langle \dot z_m,\dot z_m\rangle = -\ep^2$, we
obtain the existence of a sequence of functions $\hat A_m$ such that 
$\hat A_m \to 0$ uniformly and
$$
C_m = 
{
{-\ep^2}
\over
{\langle U(z_m),\dot z_m\rangle}
}
+
\hat A_m~.
\eqno(2.20)
$$
We show now that the functions $\langle U(z_m),\dot z_m\rangle$ are
bounded, uniformly with respect to $m \in {\N}$ and $s \in [0,1]$.
Assume by contradiction that there exists a sequence $(s_m)_{m\in{\N}}$
such that
$\langle U(z_m(s_m)),\dot z_m(s_m)\rangle \to +\infty$.
By (2.20), $C_m \to 0$ and 
$$
{
{-\ep^2}
\over
{\langle U(z_m),\dot z_m\rangle}
}
\to 0 \hbox{ uniformly }~.
$$
This means that 
$$
\vert \langle U(z_m),\dot z_m\rangle \,\vert\to +\infty 
\hbox{ uniformly }~.
\eqno(2.21)
$$
Since $U(z_m)$ is an uniformly bounded sequence of timelike vector 
fields along the curve $z_m$ and $\dot z_m$ is time like, 
$\|\dot z_m(s)\|_{\rm R} \to +\infty$ uniformly, in contradiction with
Lemma 2.6. 
Then $\langle U(z_m),\dot z_m\rangle$ is uniformly bounded with respect to
$m \in {\N}$ and $s \in [0,1]$ 
and, since $U(z_m)$ and $\dot z_m$ are timelike, there exists a positive 
constant $D$ such that
$$
\|\dot z_m(s)\|_{\rm R} \leq D, 
\forall n \in {\N}, \forall s \in [0,1]~.
\eqno(2.22)
$$
By the Ascoli--Arzel\'a Theorem, up to subsequences, 
we have that 
the sequence $(z_m)_{m\in {\N}}$ is uniformly convergent. 

\noindent 
Now,  
the sequence $(C_m)_{m \in {\N}}$ converges (up to subsequences)
to $C \in {I\!\!R}$. 
Therefore,  
the sequence 
$(\langle U(z_m),\dot z_m\rangle)_{m \in {\N}}$ is
convergent in $L^\infty$.

Now, $\langle Z_m,Z_m\rangle$ is bounded,
$z_m$  is uniformly convergent 
and $D_s Z_m = 0$. Then using (2.22) and  the Ascoli--Arzel\'a Theorem gives
that the sequence $Z_m$ has a
subsequence which is uniformly convergent.
By (2.19) there exists a subsequence 
$(\dot z_{m_k})_{m \in {\N}}$ which converges uniformly.
\cvd

\2

The manifold ${\cal L}^+_{p,\gamma,\ep}$ is only of class
$C^1$ (cf. [GMP1]). 
However, the restriction of the arrival time $\tau$ on 
${\cal L}^+_{p,\gamma,\ep}$ is of class $C^2$. 
This fact is essential to develop a Morse Theory on
${\cal L}^+_{p,\gamma,\ep}$ (in particular for the study of the behavior
of $\tau$ nearby its critical points). 

More precisely consider the $C^1$--bundle $W_\ep$ over 
the manifold ${\cal L}^+_{p,\gamma,\ep}$, whose fiber $W_\ep (z)$ is given
by the whole tangent space $T_z\Omega^{1,2}_{p,\gamma}$, 
$z \in {\cal L}^+_{p,\gamma,\ep}$, namely
$$
W_\ep (z) = 
\{(z,\zeta):z \in {\cal L}^+_{p,\gamma,\ep}, \zeta\in
T_z\Omega^{1,2}_{p,\gamma}\}~.
$$
Moreover we set
$$
W_\ep^0 = \{(z,\zeta)\in W_\ep:\zeta(1) = 0\}~.
$$
We are thinking of $W_\ep$ as a regular extension of tangent bundle 
$T{\cal L}^+_{p,\gamma,\ep}$. 
They are related by the bundle map 
$V\colon W_\ep \freccia T{\cal L}^+_{p,\gamma,\ep}$,
$$
V(z,\zeta) = (z,V_\zeta)~,
$$
where $V_\zeta$ is defined by (2.10).

\2

\noindent
{\bf Remark 2.7.}~
It is immediately checked that $V$ is a continuous map and it is a
$C^1$--map considered as a map from $W_\ep$ into itself (with image 
in $T{\cal L}^+_{p,\gamma,\ep}$). Moreover, its restriction to the tangent
bundle $T{\cal L}^+_{p,\gamma,\ep}$ is the identity map and for every 
$z \in {\cal L}^+_{p,\gamma,\ep}$, $V$ is surjective from $W_\ep^0$ to 
$T{\cal L}^+_{p,\gamma,\ep}$.

\2

\noindent
By Remark 2.7, the following proposition easily follows.

\2

\noindent
{\bf Proposition 2.8.}~
{\it The functional $\tau$ is of class $C^2$ on 
${\cal L}_{p,\gamma,\ep}^+$, in the sense that the map}
$$
(z,\zeta) \freccia \tau'(z)[V_\zeta]
$$
{\it is of class $C^1$ on $W_\ep$.}

\2

\noindent
{\bf Corollary 2.9.}~
{\it For any local chart of the manifold 
${\cal L}_{p,\gamma,\ep}^+$, the restriction of $\tau$ to the domain of
the chart is of class $C^2$.}

\2

\noindent
We prove now the timelike version of the Fermat principle for curves of
class $H^{1,1}$. 
It will be fundamental to prove Theorem 2.1.

\2

\noindent
{\bf Theorem 2.10.}~
{\it A curve $z$ is a critical point of $\tau$ on 
$\hat{\cal L}_{p,\gamma,\ep}^+$ 
in the sense that ${\rm d}\tau(z)[\zeta] = 0$ for any 
$\zeta \in T_z\hat{\cal L}_{p,\gamma,\ep}^+$
if and only if $z$ is a (smooth) geodesic.}

\1

\noindent
{\bf Proof.}~
Let $U(z)$ be the vector field along $z$ given by (2.9). 
By (2.10)-(2.11), 
$z$ is a critical point of $\tau$ if and only if for any 
vector field $W \in T_z H^{1,1}([0,1],\M)$ 
such that $W(0) = 0$, $W(1) = 0$, 
$$
\int_0^1
{
{\langle D_s W,\dot z\rangle}
\over
{\langle U(z),\dot z\rangle}
}
{\rm d}s = 0~.
\eqno(2.23)
$$
Now, assume that $z$ is a geodesic. Then $\langle U(z),\dot z\rangle$ is a
constant, since $D_s U = 0$ and $D_s\dot z =0$. Such a constant is nonzero
because $U(z)$ and $\dot z$ are both timelike.

Moreover, if $z$ is a geodesic, integration by parts gives
$$
\int_0^1 \langle D_s W,\dot z\rangle{\rm d}s = 0
$$
for all 
$W \in  H^{1,1}([0,1],T\M)$, with $W(s)\in T_{z(s)}\M$ for all
$s$, and such that $W(0) = 0$, $W(1) = 0$. 
Hence, (2.23) holds. 

Conversely, assume that (2.23) holds. Then, setting
$$
\lambda (s) = 
{
1
\over
{\langle U(z),\dot z\rangle}
}~,
$$
we have by an usual boot--strap argument that the vector field 
$\lambda(z)\dot z$ is of class $C^1$. Moreover,  $D_s(\lambda \dot z) = 0$. 
Then
$$
\hbox{ 
$\langle\lambda\dot z,\lambda\dot z\rangle = -\lambda^2\ep^2$ {\sl is
constant}, }
$$
showing that $\lambda$ is constant (and nonzero).
Then $D_s\dot z = 0$.\cvd

\2

\noindent
{\bf Remark 2.11.}~
By (2.10)-(2.11),  $z$ is a critical
point of $\tau$ if and only if $\mu(1) = 0$, for any vector field 
$W \in T_z H^{1,1}([0,1],T\M)$ along $z$, with  $W(0) = 0$, $W(1) = 0$.
Therefore, by (2.11) and Remark 2.7, 
$z$ is a critical point of $\tau$ if and only if $\zeta(1 ) = 0$, for any
$\zeta \in T_z\hat{\cal L}^+_{p,\gamma,\ep}$.

\2

We give now the statement of the well--known {\it Ekeland's variational
principle} (cf, [Ek]). It will be used in the proof of Theorem 2.1.

\2

\noindent
{\bf Theorem 2.12}~
{\it Let $(X,d)$ be a complete metric space and 
$E\colon X\freccia \R\cup \{+\infty\}$ a lower semicontinuous functional,
bounded from below, $E \not\equiv +\infty$.}

{\it Then, for any $\nu$, $\mu > 0$ and for any $u \in X$ such that}
$$
E(u) \leq \inf_X E + \mu~,
$$
{\it there exists an element $v \in X$ strictly minimizing 
the functional}
$$
E_u(w) = E(w) +{\nu \over \mu}d(u,w)~.
$$
{\it Moreover we have:}
$$
E(v) \leq E(u), 
\quad
\hbox {and } 
d(u,v)\leq \mu~.
$$

\2

\noindent
We are finally ready to prove Theorem 2.1.

\2

\noindent
{\bf Proof of Theorem 2.1.}~
Fix $q \in \M$ and choose a local chart $(U,\varphi)$ as in Remark 2.5 and including $q$.
Then we can reduce us to work on the space $V\times I$, where $V$ is a bounded
open subset of $\R^n$, $n = m - 1$, $I = ]-\lambda_0,\lambda_0[$, 
is an open interval of $\R$, 
$q = (q_0,0) \in V\times I$ and the metric $g$ satisfies (2.12).
Since $\delta$ is an integral curve of the vector field $Y$, if
$d_{\rm R}(q,\delta({{\alpha +\beta}\over 2}))$ 
is sufficiently small, we can assume that 
$$
\delta(s) = (q_\delta,s), \forall s \in ]-\lambda_0,\lambda_0[
\subset ]\alpha,\beta[~,
$$
where $
q_\delta \in V$ and $d_{\rm R}(q_0,q_\delta) \to 0$ as 
$d_{\rm R}(q,\delta({{\alpha +\beta }\over 2})) \to 0$.

If $z \in {\cal L}_{q,\delta,\ep}^+$ is a curve with values in $U$, unless
to consider the chart $\varphi(U) = V\times I$, it is $z = (x,t)$, $x(a) =
q$, $x(b) = q_\delta$ and $t$ satisfies the Cauchy problem
$$\cases{
\dot t = 
\langle {\Gamma \over \beta}(x,t),\dot x\rangle + 
\sqrt{
\langle{ \alpha \over \beta}(x,t)\dot x,\dot x\rangle +
\langle{\Gamma \over \beta}(x,t),\dot x\rangle^2 +\ep^2}
\cr
t(a) = 0\cr
}
\eqno(2.24)
$$
Moreover
$$
\tau(z) = t_x(b) = 
\int_a^b 
\langle {\Gamma \over \beta}(x,t_x),\dot x\rangle + 
\sqrt{
\langle{\alpha \over \beta}(x,t_x)\dot x,\dot x\rangle +
\langle {\Gamma \over \beta}(x,t_x),\dot x\rangle^2 +\ep^2}~ds~,
\eqno(2.25)
$$
where $t_x$ is the solution of (2.24). 
Using as a test function the chord joining $q_0$ with $q_\delta$ in the
interval $[a,b]$, we see that
$$
\inf_{{\cal L}_{q,\delta,\ep}^+}\tau \to 0 
\quad
\hbox{ as } |b-a| \to 0 
\quad
\hbox{ and }
d_{\rm R}(q_0,q_\delta) \to 0~.
\eqno (2.26)
$$
Therefore, if $|b-a|$ and $d_{\rm R}(q_0,q_\delta)$ are sufficiently
small,
$$
\hbox{ {\it 
any minimizing sequence $(z_m)_{m \in \N}$ for $\tau$ in 
${\cal L}_{q,\delta,\ep}^+$ is contained in $\varphi(U)$.} }
\eqno(2.27)
$$
The Cauchy problem (2.24) can be obviously be written as
$$\cases{
\dot t = 
\langle A(x,t),\dot x\rangle + 
\sqrt{
\langle L(x,t)\dot x,\dot x\rangle^2 +\ep^2}
\cr
t(0) = 0\cr
}
$$
where $L$ is a smooth definite operator and $A$ is a smooth vector field. 
Using the above position and the Gronwall Lemma shows that the map 
$\Phi \colon H^{1,1}([a,b],\R^n)\freccia L^1([0,1],\R)$ such that $\Phi(x)$
is the unique solution of (2.25) (whenever it is defined in all the
interval $[a,b]$) is a continuous map (cf. also [GM]).

We claim that for any $\zeta \in C^1([0,1],\R^n)$, 
$$
\Phi \hbox{ {\it is differentiable along the direction $\zeta$.} }
\eqno(2.28)
$$ 
Towards this goal consider the map
$$
G(x,t) = \dot t - 
\langle A(x,t),\dot x\rangle -
\sqrt{
\langle L(x,t)\dot x,\dot x\rangle^2 +\ep^2}~.
$$
Fix $\zeta$ of class $C^1$. 
It is $G(x,\Phi(x)) = 0$ and for any $\lambda \in \R$, 
$G(x+\lambda\zeta,\Phi(x+\lambda\zeta)) = 0$. 
Since $\zeta$ is of class $C^1$, straightforward computations shows that
there exists
$$
\lim_{\lambda \to 0}
{
{G(x+\lambda\zeta,t)-G(x,t)}
\over
\lambda
}
= 
{
{\partial G}
\over
{\partial x}
}(x,t)[\zeta]
\quad
\hbox{ {\it uniformly in $t$} }
$$
with respect to the $L^1$--norm, and
$$
{
{\partial G}
\over
{\partial x}
}
\left( x+\sigma\lambda\zeta,\Phi(x+\lambda\zeta)\right) [\zeta] 
\to {
{\partial G}
\over
{\partial x}
}
(x,\Phi(x))[\zeta] \quad
\hbox{ {\it in $L^1$} }
$$
as $\lambda \to 0$ uniformly on $\sigma \in [0,1]$.

Moreover, for any $\theta \in H^{1,1}([0,1],\R)$, 
$$\displaylines{
{
{\partial G}
\over
{\partial t}
}
(x,t)[\theta] = 
\cr
\dot\theta - 
\langle
{
{\partial A}
\over
{\partial t}
}
(x,t),\dot x\rangle\theta
-
{
1
\over 
{2\sqrt{\langle L(x,t)\dot x,\dot x\rangle +\ep^2\rangle}}
}
\left\langle 
{
{\partial L}
\over
{\partial t}
}
(x,t)\dot x,\dot x\right\rangle\theta~.\cr}
$$
This allows to show that the map 
$$
{
{\partial G}
\over
{\partial t}
}
\colon H^{1,1}([0,1],\R)\freccia L^1([0,1],\R)
$$
is invertible (the inverse can be evaluated solving a linear ordinary
differential equation) and
$$
\left[
{
{\partial G}
\over
{\partial t}
}
\left(x,\Phi(x) + \sigma(\Phi(x+\lambda\zeta)-\Phi(x))\right)
\right]^{-1} 
\to 
\left[
{
{\partial G}
\over
{\partial t}
}
(x,\Phi(x))
\right]^{-1}
$$
in $H^{1,1}([0,1],\R)$ (uniformly with respect to $\sigma$, 
because 
$\Phi(x+\lambda \zeta) \to \Phi (x)$ in $L^\infty([0,1],\R)$.

Now, since $G(x,\Phi(x)) = 0$ and 
$G(x+\lambda\zeta, \Phi(x+\lambda\zeta)) = 0$, 
applying the Lagrange Theorem we obtain:
$$\displaylines{
0 = 
{
{\partial G}
\over
{\partial x}}
(x+\sigma_1\lambda\zeta,\Phi(x+\lambda\zeta))[\lambda\zeta] 
+\cr
{
{\partial G}
\over
{\partial t}}
(x,\Phi(x) + \sigma_2(\Phi(x+\lambda\zeta)-\Phi(x)))
[\Phi(x+\lambda\zeta)-\Phi(x)]~.\cr}
$$
Dividing by $\lambda$ and passing to the limit as $\lambda \to 0$ gives
(2.28). 

Take a sequence $(\nu_m)_{m \in \N}$ of positive numbers such that 
$\nu_m \to 0$. By virtue of (2.27), for any $m \in \N$ we can choose a curve
$x_m$ with support contained in $V$ such that
$$
\tau(x_m) \leq \inf_{{\cal L}_{q,\delta,\ep}^+}\tau + \nu_m^2~.
$$
In Theorem 2.12 choose $\nu = \nu_m^2$, $\mu = \nu_m$ and $u = x_m$. 
Since $V$ is relatively compact, by (2.27) we can assume to be on a complete metric space. 
So, by applying Theorem 2.12 we find a point $y_m$ 
satisfying 
$$
\tau(y_m) \leq \tau(y_m + w) + \nu_m \|w\|_1~,
\eqno(2.29)
$$
for any $w \in H^{1,1}([a,b],V)$, 
and therefore for any $w \in C^1([a,b],V)$. 
Now by a density argument, $y_m$ can be chosen of class $C^1$. 
Then, by the arbitrariness of $w$, since $\tau$ is differentiable 
(in $H^{1,1}$) along the directions of class $C^1$, 
we deduce that
$$
|{\rm d}\tau(y_m)[\zeta]|\leq \ep_m \to 0~,
\eqno(2.30)
$$
for any $\zeta$ of class $C^1$ such that $\|\zeta\|_1 \leq 1$.

Indeed, taking $w = \lambda\zeta$ in (2.28) we have
$$
{
{\tau(y_m) - \tau(y_m + \lambda\zeta)}
\over
{|\lambda|\|\zeta\|_1}
}
=
{
{\tau(y_m) - \tau(y_m + \lambda\zeta)}
\over
{|\lambda|}
}
\leq \ep_m~,
$$
from which we deduce (2.30) sending $\lambda \to 0$ (first choosing $\lambda > 0$ and
then $\lambda < 0$).
Note that $y_m$ is a minimizing sequence (by Theorem 2.12).

Now, by the uniqueness of the related Cauchy problems, we see that 
$$
\{(\zeta,{\rm d}\Phi(y_m)\zeta):\zeta \in C^1([0,1],\R^n)\}
= 
T_{y_m} \hat{\cal L}_{q,\delta,\ep}^+ 
\cap C^1([0,1],T\M)~.
$$
Then thanks to the density of $C^1$ in $H^{1,1}$ we see that 
the sequence 
$(y_m,\Phi(y_m))_{m\in\N}$ is a minimizing sequence for $\tau$ satisfying the
assumptions of Proposition 2.4.
Then by Proposition 2.4, there exists a subsequence 
of $(y_m)_{m\in\N}$ 
convergent to a curve $y$ with respect to the $C^1$--topology.
Then $(y,\Phi(y))$ is a $C^1$--curve minimizing $\tau$ 
on $\hat {\cal L}_{q,\delta,\ep}^+ $.
Finally, by Theorem 2.10 we obtain that $(y,\Phi(y))$ is a geodesic, while
the uniqueness of the minimizer comes from the local invertibility of the
exponential map.\cvd

\2

\noindent
{\bf Remark 2.14}~
Working in local coordinates shows immediately that for any fixed
neighborhood ${\cal U}_q$ of $q$, there exists a positive number $\rho_q$
such that the minimal geodesic for $\tau$ on 
${\cal L}_{q,\delta,\ep}^+([a,b])$ is in ${\cal U}_q$ if
$$
d_{\rm R}\left (q,\delta\left(
{{\alpha+\beta}\over 2}\right)\right)
\leq \rho(q)
\quad
\hbox{ {\it and} }
|b-a|\leq \rho(q)
~.
$$

\vfil
\eject

%

\noindent
{\bf 3. A shortening method for $\tau$ on ${\cal L}_{q,\gamma,\ep}^+$.}

\1

\noindent

In this section we shall introduce a 
shortening flow 
for the functional $\tau(z)$. Such a flow
will be used to get the deformations for the
sublevels of $\tau$ 
(needed to develop a Morse Theory), when we are far from the critical points
of $\tau$, i.e. timelike geodesics.

To construct the shortening flow we shall use the same ideas as in
[Mi],  adapting them to our case. 
Note that here we can not use the same finite dimensional approach 
nearby critical curves (used in [Mi] for Riemannian geodesics) 
because we are not working with fixed points boundary conditions.

The shortening procedure, which is illustrated
by a five pictures appearing at the end of the paper, is constructed
in the following way. 

Fix $c > \inf\{\tau(z), z \in {\cal L}^+_{p,\gamma,\ep} \}$
and consider $D(c)$ as in Lemma 2.6. 
Let $K_{c}$ be a compact subset of $\M$ including all the curves 
$z \in {\cal L}_{q,\gamma,\ep}^+$ such that $\tau(z) \leq c.$

Let  $\rho_*(c) > 0$ be such that 
Theorem 2.1 holds with $\rho_(q)$ replaced by $\rho_*(c)$ for any  
$q \in K_c$. Take  $N = N(c)$ such that
$$
{1 \over N} \leq \rho_\ast (c)\die,
\quad
{
{D(c)}
\over
N
}
\leq \rho_*(c)\die .
$$
Choose a partition $\{0 = s_0 < s_1 \dots s_{N-1} < s_N =1\}$ of $[0,1]$
such that for any $i \in \{1, \dots N\}$,
$$
s_{i} - s_{i-1} = {1\over N}\die .
$$
For any $z \in \tau^c\cap {\cal L}^+_{p,\gamma,\ep}$, 
choose $N+1$ points
$z_0,z_1, \dots z_{N}$ on $z([0,1])$ such that $z(0) = p$, $z_N = z(1)$ and
$d_R(z_i,z_{i-1}) = l(z)/N$, for any $i \in \{1, \dots N\}$, where $l(z)$
denotes the length of $z$ with respect to the Riemannian structure (2.3)
(see Figure~1).

Denote by $\gamma_i$ ($i = 1,\dots,N)$ the maximal integral
curve of $W$ such that $\gamma_i(0) = z_i$ (see Figure~2).
Observe that $\gamma_N(s)=\gamma(s+\tau(z))$ for all $s$.

Let $w_1$ be the geodesic minimizing $\tau$ on
${\cal L}^+_{p,\gamma_1,\ep}([s_0,s_1])$ 
(recall that $z_0 = p$ and $s_0=0$),
$w_2$ the lightlike geodesic minimizing $\tau$ on
${\cal L}^+_{w_1(s_1),\gamma_2,\ep}([s_1,s_2])$, and so on 
(see Figure~3).

In Figures 3, 4 and 5 the points $w_i(s_i)$ are denoted by
$\overline w_i$.

\noindent
Note that the number $N$ can be chosen large enough so that
$d_R(w_i(s_i),z_{i+1}) \leq \rho_*(c)$, for any $i=1,\dots,N-1$ 
and for any $z\in\tau^c$.

\2

\noindent
{\bf Remark 3.1.}~
Let $K=K(c)$ be a compact subset of $\M$ containing the images of the
curves of the curves $z \in {\cal L}^+_{p,\gamma,\ep}$, with $\tau(z)\leq c$. 
By compactness, $K(c)$ can be covered by a finite family
$(U_j)$ as in Remark 2.5, and the Lorentzian metric $g$ is described by
(2.12). 
 
Moreover, $N$ can be chosen so large that
$z([s_{i-1},s_i])$ and the minimizer of $\tau$ on
${\cal L}^+_{w_{i-1}(s_{i-1}),\gamma_i,\ep}([s_{i-1},s_i])$
are contained in some $U_j$.

\2

\noindent
With the notation of Remark 2.5,  
for any  future pointing curve $z$ with image contained in some
$U_j$,  the condition
$\langle\dot z,\dot z\rangle = - \ep^{2}$ holds if and only if
$$
\dot t =
\langle {{\Gamma_j}\over{\beta_j}}(x,t), \dot x \rangle_0
+
\sqrt{
\langle{{\alpha_j}\over{\beta_j}}(x,t),\dot x,\dot x\rangle_0
+
\langle{{\Gamma_j}\over{\beta_j}}(x,t),\dot x\rangle_0^2
+ \ep_2}
\eqno(3.1)
$$
Moreover, any $\gamma_i$ is an integral curve of $W$, so, in $U_j$, it
has the form $s \longmapsto (x_j,t{_j}+s)$, if $z_j=(x_j,t_j)$.

Note that
${\cal L}^+_{p,\gamma_1,\ep}([s_0,s_1])$
is nonempty, since it contains the restriction $z_{|[s_0,s_1]}$.

Now, using elementary comparison theorems for ordinary differential equations
allow to deduce that also any
space
${\cal L}^+_{w_{i-1}(s_{i-1}),\gamma_{i},\ep}([s_{i-1},s_i])$
is nonempty for
any $i \in \{2,\dots N\}$.

Note also that, if $\eta_1$ is the curve defined by setting
$\eta_1 ([s_{i-1},s_i]) = w_i$, 
then $\tau(\eta_1) \leq \tau(z)\le c$ (always by
comparison theorems in O.D.E.).
In particular $\eta_1([0,1])$ is contained in $K(c)$.

\2

\noindent
{\bf Remark 3.2}~
A second curve $\eta_2$ will be constructed
in the following way starting from $\eta_1$.
On any minimizer $w_i$ ($i=1,\dots,N$) consider the point $m_{i}$
such that
$d(w_i(s_{i-1}), m_i) = d(m_i,w(s_i))$.

For $i =1,\dots,N$, 
we denote by $\lambda_i$ the maximal integral curve of $W$ such that
$\lambda_i(0) = m_i$; moreover, we set $\lambda_{N+1}(s) =
\gamma(s+\tau(\eta_1))$ (see Figure~4).

Consider now the following subdivision of the interval $[0,1]$. Let
$\sigma_0=0$, $\sigma_1={1\over 2N}$, 
$\sigma_j={2j-1\over 2N}$ for $j=2,\ldots,N$,
and $\sigma_{N+1}=1$.

Denote by $u_1$ the minimizer of $\tau$ on 
${\cal L}^+_{p,\lambda_1,\ep}([\sigma_0,\sigma_1])$,
by $u_2$ the minimizer of $\tau$ on
${\cal L}^+_{u_1(\sigma_1),\lambda_2,\ep}([\sigma_1,\sigma_2])$ 
and so, inductively,
we denote by $u_j$ the minimizer of $\tau$ in 
${\cal L}_{u_{j-1}(\sigma_{j-1}),
\lambda_{j},\ep}([\sigma_{j-1},\sigma_j],)$, $j=2,\ldots,N+1$.

Finally, (see Figure~5) we denote by
$\eta_2$ the curve such that
$\eta_2\vert_{[\sigma_{j-1},\sigma_j]}=u_j$.

Using again comparison theorems in ordinary differential equations
one proves that
$\tau(\eta_2) \leq \tau(\eta_1)$.

\2

\noindent
The continuous flow $\eta(\sigma,z)$ can be constructed as follows.
Fix $\sigma \in [0,1]$ and consider for instance the interval
$[s_0,s_1]$.
We choose $\eta(\sigma,z)_{|[s_0,s_1]}$ as follows.
Set $p = (x_0,0)$ and $\gamma_1(s) = (x_1,t_1+s)$
(in some neighborhood $U_j$ as in Remark 3.1).
Since $z(s) = (x(s),t(s))$, the curve $x(s)$ joins $x_0$ with $x_1$.

Let $y(\sigma)$ be the minimizer of the functional
$$\displaylines{
y \longmapsto
\int_{s_0}^{\sigma s_1}
\langle{{\Gamma_i}\over{\beta_i}}(y,t_y),\dot y\rangle_0{\rm d}s
+
\cr
\hfill
\int_{s_0}^{\sigma s_1}
\sqrt{
\langle{{\alpha_i}\over{\beta_i}}(y,t_y),\dot y,\dot y\rangle_0
+
\langle
{{\Gamma_i}\over{\beta_i}}(y,t_y),\dot y\rangle_0^2
}
~{\rm d}s
~,\hfill\llap{(3.2)}
\cr}
$$
with boundary conditions $y(0) = x_0$ and $y(\sigma s_1) = x(\sigma
s_1)$,
where $t_y$ is the solution of (3.2) with 
$t_y(0) = 0$ in the interval $[0,{\sigma s_1}]$.

Denote by $\hat y(\sigma)$ the extension of $y(\sigma)$ to $[s_0,s_1]$
taking $\hat y(s) = x(s)$ for $s \in [\sigma s_1,s_1]$.
Finally, denote by $\hat t_y$ the corresponding solution of (3.1) in the
interval $[s_0,s_1]$.
The curve $(\hat y(\sigma),\hat t_y(\sigma))$ will be $\eta(\sigma,z)$ in the
interval $[s_0,s_1]$.
In the same way we can construct $\eta(\sigma,z)$ on the other intervals
$[s_{i-1},s_i]$. Note that, by construction,  $\eta(1,z)=\eta_1$.
Similarly, we can extend the flow $\eta$ to a map defined
on $[0,2]\times\tau^c$ in such a way that $\eta(2,z)=\eta_2$.

Now, we iterate the shortening argument above,  
replacing the original curve $z$ with
the curve $\eta_2$.
Successively we apply the above construction, starting from $\eta_2$.
By induction we
obtain a flow $\eta(\sigma,z)$, defined on $\R^+\times\tau^c$.
Note that $\tau(\eta(\sigma,z))\le\tau(z)$ for any $\sigma$ and for any $z$.

Suppose that $\tau(\eta_1)=\tau(\eta_2)$ 
and consider the situation is a single interval $[\sigma_j,\sigma_{j+1}]$.
Since $\tau(\eta_1)=\tau(\eta_2)$ 
simple comparison theorems in O.D.E. show that 
$\eta_1$ is a minimizer on the interval $[\sigma_j,\sigma_{j+1}]$.
Suppose that it consists of two (nonconstant) lightlike geodesics. 
If it is not a light like geodesic, 
by the above construction it has a discontinuity at 
$s_{j+1} = 
{
{\sigma_{j+1} + \sigma_{j}}
\over
{2}
}
$.
Denote by $U_{\eta_{1}}$ the parallel transport of 
$\dot \gamma(\tau(\eta_{1}))$ 
along the curve $\eta_1$. 
Since $\eta_1$ is a minimizer, by (2.23) it is 
$$
\int_{\sigma_j}^{\sigma_{j+1}}
{\langle D_{s}V,{\dot \eta_1 \rangle} 
\over 
{\langle U_{\eta_{1}},\dot \eta_1 \rangle}} ds = 0
$$
for any $C^{\infty}$-vector field along $\eta_1$ such that $V(0)=0, V(1)=0$.
In particular 
${\dot \eta_1} \over {\langle U_{\eta_{1}},\dot \eta_1 \rangle}$ is a
$C^1$ curve
and also
$$
{
{-\ep^2}
\over
{\langle U_{\eta_1},\dot \eta_1\rangle^2}
}
$$
is of class $C^1$. 
And this implies that $\eta_1$ is of class $C^1$, because 
$\langle U_{\eta_1},\dot \eta_1\rangle$ never changes its sign.

Then, whenever we are far from critical 
points of $\tau$ on ${\cal L}^+_{p,\gamma,\ep}$, $\tau(\eta_2) < \tau(\eta_1)$.  

Finally compactness arguments similar to the ones used for 
the shortening method for Riemannian geodesics (cf. [Mi]), allows to obtain  
the analogous of the classical deformation results (cf e.g. [MW,St])  
for the functional $\tau$ on 
${\cal L}^+_{p,\gamma,\ep}$. 

For any $d \in \R$ set $\tau^{d} = \{z \in {\cal L}^+_{p,\gamma,\ep} : \tau(z) \leq d \}.$

\2

\noindent
{\bf Proposition 3.3.}~
{\it
Let $c$ be a regular value for $\tau$ on
${\cal L}^+_{p,\gamma,\ep}$ (namely $\tau^{-1}(\{c\})$ does not 
contain geodesics). 

Then, there exists a positive number $\delta = \delta (c)$ and a continuous
map
$H \in C^0([0,1]\times \tau^{c+\delta},\tau^{c+\delta})$, such that:}

\smallskip

\item{(a)} $H(0,z) = z$, {\it for every $z \in  \tau^{c+\delta}$;}
\smallskip
\item{(b)} $H(1,\tau^{c+\delta}) \subseteq \tau^{c-\delta}$;\smallskip

\item{(c)} $H(\sigma,z) \in \tau^{c-\delta}$,
{\it for any $\sigma\in [0,1]$ and $z \in \tau^{c-\delta}$}.

\2

\noindent
{\bf Proposition 3.4.}~
{\it 
Let $Z_c$ be the set of  the timelike geodesics on 
${\tau}^{-1}(\{c\}) \cap {\cal L}^+_{p,\gamma,\ep}$.
Then for any open neighborhood $\cal U$ of $Z_c$, there exists a positive number
$\delta = \delta({\cal U},c)$ and a homotopy
$H \in C^0([0,1]\times \tau^{c+\delta},\tau^{c+\delta})$, such that}
\smallskip

\item{(a)} $H(0,z) = z$, {\it for any $z \in  \tau^{c+\delta}$.}
\smallskip

\item{(b)}
$H(1,\tau^{c+\delta} \setminus {\cal U}) \subset \tau^{c-\delta}$;
\smallskip

\item{(c)} $H(\sigma,z) \in \tau^{c-\delta}$,
{\it for every $\sigma\in [0,1]$ and $z \in \tau^{c-\delta}$}.

\2

\noindent
{\bf Remark 3.5.}\die
There are two main differences
between the shortening method described above
and the classical shortening method for Riemannian
geodesics.  In our case, we locally minimize a functional which is
is not given in an integral form.
Secondly, we minimize the functional in the space of curves joining
a point with a curve, and not two fixed points.

\2

\noindent
{\bf Remark 3.6.} 
The flow used in proving 
Propositions 3.3--3.4 are just what 
we need for a Ljusternik--Schnirelmann theory. 
Then, without using the nondegeneracy
assumption of Theorem 1.3 we can obtained the existence of at last 
$cat ({\cal L}^+_{p,\gamma,\ep})$ future pointing timelike geodesics in
${\cal L}^+_{p,\gamma,\ep}$. 
(Here cat $X$ denotes the minimal number of
contractible subsets of $X$ covering it). 
Moreover if 
$cat ({\cal L}^+_{p,\gamma,\ep})=+\infty$ 
there is a sequence $z_n$ of future
pointing timelike geodesics in ${\cal L}^+_{p,\gamma,\ep}$ such that
$\tau (z_n)\to +\infty$. (Recall that we are assuming that $\gamma$ is defined on $\R$).

\vfil
\eject
%
%
%
%

\noindent
{\bf 4. The index Theorem and the Morse Relations on 
${\cal L}^+_{p,\gamma,\ep}$}

\1

\noindent
In this section we shall prove the Morse Relations on 
${\cal L}^+_{p,\gamma,\ep}$ and the second part of Theorem 1.2, namely

\2

\noindent
{\bf Theorem 4.1}~
{\it Let $z$ be a geodesic in ${\cal L}^+_{p,\gamma,\ep}$ such that $z(1)$
is nonconjugate to $p$ along $z$.
Then:}
$$
\mu(z) = m(z,\tau)~,
$$
{\it where $\mu(z)$ is the geometric index of $z$ and $m(z,\tau)$ is the
Morse index of $z$ considered as a critical point of $\tau$ on 
${\cal L}^+_{p,\gamma,\ep}$.}

\2

\noindent
In order to prove Theorem 4.1, we first need to evaluate the Hessian of
$\tau$ at $z$, 
$$
H_\tau(z)[\zeta,\zeta] = 
{
{d^2}
\over
{d\sigma^2}
}
\left(
\tau(\eta(\sigma,\cdot))
\right)_{\sigma = 0}~,
$$
where $\zeta \in T_z{\cal L}^+_{p,\gamma,\ep}$ and 
$\eta\colon ]-\sigma_0,\sigma_0[\freccia {\cal L}^+_{p,\gamma,\ep}$ is a
variation of $z$ with variational vector field $\zeta$, that is
$$\displaylines{
\eta(0,s) = z(s), \hbox{ for any $s \in [0,1]$ };\cr
\eta_\sigma (0,s) = \zeta(s), \hbox{ for any $s \in [0,1]$ }.\cr}
$$
Here $\eta_\sigma$ denotes the partial derivative with respect to
$\sigma$. 

\2

\noindent
{\bf Proposition 4.2.}~
{\it In the notation above, for all 
$\zeta\in T_z{\cal L}^+_{p,\gamma,\ep}$, it is:}
$$
H_\tau(z)[\zeta,\zeta] =
{
{-1}
\over
{\langle \dot\gamma(\tau(z)),\dot z(1)\rangle}
}
\int_0^1
\left(
\langle D_s\zeta,D_s\zeta\rangle - 
\langle R(\zeta,\dot z)\dot z,\zeta\rangle
\right)~{\rm d}s~.
\eqno(4.1)
$$

\1

\noindent
{\bf Proof.}~
Since $\eta(\sigma,\cdot) \in {\cal L}^+_{p,\gamma,\ep}$ for any $\sigma$,
we have
$$
\langle\eta_s(s,\sigma),\eta_s(s,\sigma)\rangle = -\ep^2, \quad\quad
\hbox{ for any $s$ and for any $\sigma$ }~.
$$
Here $\eta_s$ denotes the partial derivative of $\eta$ with respect to
$s$. 
Since $z$ is of class $C^2$, it suffices to prove (4.1) whenever $\zeta$
(and therefore $\eta$) is of class $C^2$ and apply standard density
arguments. We have:
$$
{
{\partial}
\over
{\partial\sigma}
}
\left(
\int_0^1 \langle\eta_s,\eta_s\rangle {\rm d}s
\right) = 0
$$
and therefore
$$\displaylines{
0 = \int_0^1 \langle D_\sigma\eta_s,\eta_s\rangle {\rm d}s = 
\int_0^1 \langle D_s\eta_\sigma,\eta_s\rangle {\rm d}s = \cr
\hfill
\langle \eta_\sigma(\sigma,1),\eta_s(\sigma,1)\rangle - 
\langle \eta_\sigma(\sigma,0),\eta_s(\sigma,0)\rangle - 
\int_0^1 \langle \eta_\sigma,D_s\eta_s\rangle {\rm d}s~.
\hfill\llap{(4.2)}\cr}
$$
Now, since $\gamma(\tau(\eta(\sigma,\cdot))) = \eta(\sigma,1)$, we have
$$
\dot\gamma(\tau(\eta(\sigma,\cdot)))
{
{{\rm d}\tau}
\over
{{\rm d}\sigma}
}
\left(
\eta(\sigma,\cdot)
\right) 
= 
\eta_\sigma(\sigma,1)~,
$$
therefore, since $\eta_\sigma(\sigma,0) = 0$ for any $\sigma$, by (4.2) we
have:
$$\displaylines{
{
{{\rm d}\tau}
\over
{{\rm d}\sigma}
}
\left(
\eta(\sigma,\cdot)
\right) 
=
{
{\langle \eta_\sigma(\sigma,1),\eta_s(\sigma,1)\rangle}
\over
{\langle \dot\gamma(\tau(\eta(\sigma,\cdot))),\eta_s(\sigma,1)\rangle}
}\cr
= 
{1
\over
{\langle \dot\gamma(\tau(\eta(\sigma,\cdot))),\eta_s(\sigma,1)\rangle}
}
\int_0^1 \langle \eta_\sigma,D_s\eta_s\rangle~{\rm d}s~.
\cr}
$$
Note that 
$\langle \dot\gamma(\tau(\eta(\sigma,\cdot))),\eta_s(\sigma,1)\rangle
\not=0$, because both 
$\dot\gamma(\tau(\eta(\sigma,\cdot)))$ and 
$\eta_s(\sigma,1)$ are timelike vectors.

Then, since $D_s \dot z = 0$, we get
$$\displaylines{
{
{{\rm d}^2\tau}
\over
{{\rm d}\sigma^2}
}
\left(
\eta(\sigma,\cdot)
\right)_{|\sigma = 0 } = \cr
{
{{\rm d}}
\over
{{\rm d}\sigma}
}
\left(
{1
\over
{\langle \dot\gamma(\tau(\eta(\sigma,\cdot))),\eta_s(\sigma,1)\rangle}
}
\right)
\int_0^1 \langle \zeta,D_s\eta_s\rangle~{\rm d}s + \cr
{1
\over
{\langle \dot\gamma(\tau(\eta(\sigma,\cdot))),\eta_s(\sigma,1)\rangle}
}
{
{{\rm d}}
\over
{{\rm d}\sigma}
}
\left(
\int_0^1 \langle \eta_\sigma,D_s\eta_s\rangle~{\rm d}s
\right)_{\sigma = 0} = \cr
{1
\over
{\langle \dot\gamma(\tau(\eta(\sigma,\cdot))),\eta_s(\sigma,1)\rangle}
}
\left(
\int_0^1 
(
\langle D_\sigma\eta_\sigma,D_s\eta_s\rangle +
\langle\eta_\sigma,D_\sigma D_s\eta_s\rangle
)~{\rm d}s
\right)
_{\sigma = 0} =\cr
{1
\over
{\langle \dot\gamma(\tau(\eta(\sigma,\cdot))),\eta_s(\sigma,1)\rangle}
}
\left(
\int_0^1 
\langle\eta_\sigma,D_\sigma D_s\eta_s\rangle~{\rm d}s
\right)_{\sigma = 0}~.\cr}
$$
Since 
$D_\sigma D_s\eta_s = D_s D_\sigma\eta_s + R(\eta_\sigma,\eta_s)\eta_s$
(cf. [BEE]), 
we have:
$$
H^\tau(z)[\zeta,\zeta] = 
{1
\over
{\langle \dot\gamma(\tau(\eta(\sigma,\cdot))),\eta_s(\sigma,1)\rangle}
}
\int_0^1
\left(
\langle\eta_\sigma,D_s D_\sigma\eta_s + R(\eta_\sigma\eta_s)
\eta_s\rangle~{\rm d}s
\right)_{\sigma = 0}
$$
$$\displaylines{
= 
{1
\over
{\langle \dot\gamma(\tau(\eta(\sigma,\cdot))),\eta_s(\sigma,1)\rangle}
}
\left(
\langle\eta_\sigma(\sigma,1),D_\sigma\eta_s(\sigma,1)\rangle - 
\langle\eta_\sigma(\sigma,0),D_\sigma\eta_s(\sigma,0)\rangle
\right)_{\sigma = 0}\cr
+
{1
\over
{\langle \dot\gamma(\tau(\eta(\sigma,\cdot))),\eta_s(\sigma,1)\rangle}
}
\left[
\left(
-\int_0^1 
\langle D_s\eta_\sigma,D_\sigma\eta_s\rangle~ds +
\int_0^1 \langle R(\eta_\sigma,\eta_s)\eta_s,\eta_\sigma\rangle~ds
\right)
\right]_{\sigma = 0}
\cr}
$$
$$\displaylines{
= 
{1
\over
{\langle \dot\gamma(\tau(\eta(\sigma,\cdot))),\eta_s(\sigma,1)\rangle}
}\cdot
\cr
\left(
\langle\zeta(1),D_{\zeta(1)}\dot z(1)\rangle - 
\langle\zeta(0),D_{\zeta(0)}\dot z(0)\rangle
-\int_0^1 
\langle D_s\zeta,D_s\zeta\rangle~ds +
\int_0^1 \langle R(\zeta,\dot z)\dot z,\zeta\rangle~ds
\right)\cr}
$$
Finally, $\zeta(0) = 0$ and by Remark 2.11, $\zeta(1) = 0$.
\cvd

\2

Let $z$ be a geodesic in ${\cal L}^+_{p,\gamma,\ep}$. 
For any $\theta \in ]0,1]$, set 
$$\displaylines{
A_\theta = 
\{\zeta \in H^{1,2}([0,\theta],T\M): \zeta(s) \in T_{z(s)}\M 
\hbox{ for any $s \in [0,\theta]$ }, \cr
\hfill
\langle D_s\zeta,\dot z\rangle = 0 
\hbox{ a.e. } ,
\zeta(0) = 0, \zeta(\theta)) = 0 \}.
\cr}
$$
We consider the bilinear form on $A_\theta$ given by
$$
J_\theta(z)[\zeta,\zeta] =
\int_0^\theta 
\left(
\langle D_s\zeta,D_s\zeta\rangle - 
\langle R(\zeta,\dot z)\dot z,\zeta\rangle\right)
~ds
~.
\eqno(4.3)
$$

\2

\noindent
{\bf Lemma 4.3.}~
{\it Let $\zeta_{0} \in A_{\theta}$. In the above notations, 
$J_\theta(z)[\zeta_0,\cdot] = 0$ in $A_\theta$, if and only if $\zeta_0$
solves {\rm (1.2)} in $[0,\theta]$.}

\1

\noindent
{\bf Proof.}~
Let $V \in C^\infty_0([0,\theta],T\M)$ such that 
$V(s) \in T_{z(s)}\M$, for any $s \in [0,\theta]$. 
Since $z$ is a geodesic, we can describe all the elements of $A_\theta$ by
$$
\zeta = V + 
{
{\langle V, \dot z\rangle}
\over 
{\ep^2}
}
\dot z~.
$$
Indeed, $\zeta(0) = 0$, $\zeta(\theta) = 0$ and 
$\langle D_s\zeta,\dot z\rangle = 0$, because 
$\langle\dot z,\dot z\rangle = -\ep^2$ and $D_s\dot z = 0$.

Then, $J_\theta (z)[\zeta_0,\zeta] = 0$ for any 
$\zeta\in A_\theta$ if and only if
$$
\int_0^\theta
\left(
\langle D_s\zeta_0,
D_s V + {{\langle D_s V,\dot z\rangle\dot z}\over{\ep^2}}\dot z\rangle 
-\langle R(\zeta_0,\dot z)\dot z,
V + {{\langle V,\dot z\rangle\dot z}\over{\ep^2}}\rangle
\right)
~ds = 0
~,
$$
for any $V \in C^\infty_0([0,1],\M)$ such that 
$V(s) \in T_{z(s)}\M$, for any $s \in [0,\theta]$.

But $\langle D_s\zeta_0,\dot z\rangle = 0$, because $\zeta_0 \in A_\theta$
and 
$\langle R(\zeta_0,\dot z)\dot z,\dot z\rangle = 0$ by well known properties
of the Riemann tensor. 
Therefore $J_\theta(z)[\zeta_0,\zeta] = 0$ for any $\zeta \in A_\theta$,
if and only if
$$
\int_0^\theta
\left(
\langle D_s\zeta_0,D_s V\rangle 
-\langle R(\zeta_0,\dot z)\dot z,
V\rangle
\right)
~ds = 0
~,
$$
for any $V \in C^\infty_0([0,1],\M)$ with  
$V(s) \in T_{z(s)}\M$ for any $s \in [0,\theta]$.
Then, an integration by parts completes the proof.
\cvd

\2

\noindent
We are finally ready to prove Theorem 4.1.

\2

\noindent
{\bf Proof of Theorem 4.1.}~
Recalling (2.4), since $\dot z$ is a timelike vector field, a simple
compactness argument shows the existence of $\nu = \nu(z) > 0$ such that
$$
\langle w,w\rangle \geq \nu (z)\langle w,w\rangle_{\rm (R)}~,
$$
for any vector field $w$ along $z$, such that 
$\langle w(s),\dot z(s)\rangle = 0$ for any $s$. 
Moreover, since $\gamma$ is an integral curve of $Y$ and $\dot z(1)$ is
future pointing, 
$\langle\dot\gamma(\tau(z)),\dot z(1)\rangle < 0$. 
Therefore, by (4.1), for any $\theta \in ]0,1]$ the linear operator
associated to the bilinear form $J_\theta$ is a compact perturbation of
the identity operator if we equip $A_\theta$ with the natural Riemannian
structure given by
$$
\int_0^\theta
\langle D^{\rm (R)}_s\zeta,D^{\rm (R)}_s\zeta\rangle_{\rm (R)}~{\rm d}s~.
$$
Then we can use the methods of Milnor in [Mi] (cf. also [Ma]) and Lemma
4.3 to conclude the proof.
\cvd

\2

\noindent
Now we can prove the classical Morse Relations on the sublevels of $\tau$ on 
${\cal L}^+_{p,\gamma,\ep}$ 
They can be stated in the following way. 
For any $b \in {I\!\!R}\cup \{+ \infty\}$ set:
$$\displaylines{
{\cal G}^{+,b}_{p,\gamma,\ep} = 
\{z\in C^2([0,1],\M): 
\hbox{ $z$ is a future pointing geodesic such that: } \cr
z(0) = p, z(1) \in {\gamma}(\R), 
{\langle {\dot z,\dot z} \rangle} \equiv {-\ep^2},
\tau(z) \leq b \}
\cr}
$$

\2

\noindent
{\bf Theorem 4.4.}~
{\it Assume that $\lor$ is strongly causal 
and that assumptions {\rm 1)---3)} of {\rm Theorem 1.3} hold true.

Then, for any field ${\cal K}$ and for any regular value $b$ of $\tau$ on 
${\cal L}^+_{p,\gamma,\ep}$, $b \in ]\inf\tau,+\infty]$, 
there exists a formal series
${\cal S}(\lambda)$ with non negative integer coefficients (possibly 
$+ \infty$  if $b = +\infty$) such that:}
$$
\sum_{z \in {\cal G}^{+,b}_{p,\gamma,\ep}}
\lambda^{\mu(z)} = 
{\cal P}_\lambda(\tau^b,{\cal K}) + (1+\lambda){\cal S}(\lambda)~,
\eqno(4.4)
$$
{\it where 
${\cal P}_\lambda(\tau^b,{\cal K})$ is the Poincar\'e polynomial of
$\tau^b$ with coefficients in ${\cal K}$.}

\1

\noindent
{\bf Proof.}~
By Lemma 4.3 and assumption 2) of Theorem 1.3, any critical point of
$\tau$ on ${\cal L}^+_{p,\gamma,\ep}$ is nondegenerate 
(and therefore isolated). 
Moreover, using the geodesic
equation, it is not difficult to prove that, for every $b\in{I\!\!R}$,
the set 
${\cal G}^{+,b}_{p,\gamma,\ep}$ is compact with respect to the
$C^2$--topology. Hence, for all $b\in{I\!\!R}$,
the set ${\cal G}^{+,b}_{p,\gamma,\ep}$ is finite.

By the deformation results of Propositions 3.3--3.4, since  
$\tau$ is of class $C^2$ on the Hilbert manifold 
${\cal L}^+_{p,\gamma,\ep}$ 
we can apply the classical Morse Theory (cf. [C,MW]) to describe 
the topology nearby the geodesics, obtaining the classical Morse Relations
$$
\sum_{z \in {\cal G}^{+,b}_{p,\gamma,\ep}}
\lambda^{m(z,\tau)} = 
{\cal P}_\lambda(\tau^b,{\cal K}) + (1+\lambda){\cal S}(\lambda)~.
$$
Here $m(z,\tau)$ denotes the Morse index of the critical points $z$ 
for the functional 
$\tau$ in the Hilbert manifold ${\cal L}^+_{p,\gamma,\ep}$.  
Finally, thanks to Theorem 4.1, the Morse Relations (4.4) follow.
\cvd
 
\2

\noindent
{\bf Proof of Theorem 1.3.}~
By Theorem 4.4, setting $b = +\infty$ we have 
$$
\sum_{z \in {\cal G}^+_{p,\gamma,\ep}}
\lambda^{\mu(z)} = 
{\cal P}_\lambda({\cal L}^+_{p,\gamma,\ep},{\cal K}) + 
(1+\lambda){\cal S}(\lambda)~,
$$
obtaining the proof.
\cvd

\vfil
\eject


\noindent
{\bf 5. Some relations between ${\cal L}^+_{p,\gamma}$ and 
${\cal L}^+_{p,\gamma,\ep}$}

\1

\noindent
In this section we will discuss the method of approximation of
the space ${\cal L}^+_{p,\gamma}$ 
with the regular manifolds 
${\cal  L}_{p,\gamma,\ep}^+$, pointing the results needed to obtain
the Morse Relations on 
${\cal L}^+_{p,\gamma}$ as limit of the Morse Relations on 
${\cal  L}_{p,\gamma,\ep}^+$.
The first result, which is stated in the following proposition, is concerned
with
the existence of transition functions between  
${\cal  L}_{p,\gamma}^+$
and ${\cal  L}_{p,\gamma,\ep}^+$.

\2

\noindent
{\bf Proposition 5.1.}~
{\it 
Suppose that 
$\tau$ is pseudo--coercive in ${\cal  L}_{p,\gamma}^+$. 
Then, for any $c > \inf \tau$, there exists a
positive number $\epsilon_0 = \ep_0(c) > 0$ such that, for every
$\ep\in ]0,\ep_0]$ there exist two injective maps:}
$$\displaylines{
\phi_\ep \colon \tau^c\cap {\cal  L}_{p,\gamma}^+ \freccia
 {\cal  L}_{p,\gamma,\ep}^+~,\cr
\psi_\ep \colon 
{\cal  L}_{p,\gamma,\ep}^+ \freccia {\cal  L}_{p,\gamma}^+~,
\cr}
$$
{\it such that:}

\item{(1)}
{\it 
$\phi_\ep$ and $\psi_\ep$ are continuous with respect to the $H^{1,1}$-norm;}

\item{(2)}
{\it for every $z\in \tau^c\cap {\cal  L}_{p,\gamma}^+$ it is
$\psi_\ep(\phi_\ep (z))=z$;}

\item{(3)}
{\it for every $z\in {\cal  L}_{p,\gamma,\ep}^+$ 
such that $\tau(\psi_\ep (z))\leq~c$, it is
$\phi_\ep(\psi_\ep (z))=z$;}

\item{(4)}
{\it 
if $\ep_1 < \ep_2$, then $\tau(\phi_{\ep_1}(z)) \leq 
\tau(\phi_{\ep_2}(z))$ and $\tau(\psi_{\epsilon_1}(z))\ge
\tau(\psi_{\epsilon_2}(z))$;}

\item{(5)}
{\it $\tau(\phi_{\epsilon}(z)) \geq \tau(z)$ and
$\tau(\psi_{\epsilon}(z)) \leq \tau(z)$;}

\item{(6)}
{\it there exists a positive constant $M=M(c)$ such that
${\rm d}_2(\phi_\ep (z),z)\leq M \cdot\ep$ for every $z\in
\tau^c\cap {\cal L}_{p,\gamma}^+$, where $d_2$ is the metric induced by
the Hilbert structure {\rm (2.4)}}.

\1

\noindent
{\bf Proof.}~
We fix $c$ and we find a compact subset $K$ such that
the support of every
$z\in \tau^c\cap {\cal  L}_{p,\gamma}^+$ lies  in $K$.  
Let $\delta$ be a positive number such that the flow 
$\Phi(s,q)$ of the vector field $Y$ is defined on
$[-\delta,\delta]\times K$.
By definition, the curve $\eta_q(s)=\Phi(s,q) $ is
the maximal solution of the Cauchy problem:
$$\cases{
\dot\eta = Y(\eta),\cr
\eta(0) = q.\cr}
$$
For $z\in \tau^c\cap {\cal  L}_{p,\gamma}^+$, we define
$$
z_\ep (s) = \phi_\ep (z)(s) =
\Phi(\sigma_{z,\ep}(s),z(s)),
$$
for some function
$\sigma_{z,\ep}(s)=\sigma(s)$ on $[0,1]$ and with values
in
$[0,\delta)$, to be determined in such a way that
$$
\sigma_{z,\ep}(0) =0~,
$$
(which means that $z_\varepsilon(0)=p$),
$$
\langle \dot z_\ep,Y(z_\ep)\rangle < 0
\eqno(5.1)
$$
and 
$$
\langle \dot z_\ep,\dot z_\ep\rangle = -\ep^2~.
$$
Observe that any such curve automatically satisfies
$z_\ep (1)\in\gamma({I\!\!R})$,
since $\gamma$ is an integral curve of $Y$ and
$\Phi(0,z(1))=z(1)\in \gamma({I\!\!R})$. 

We compute $\dot z_\ep$ as follows:
$$
\dot z_\ep = \Phi_q[\dot z]+\Phi_\sigma[\dot\sigma] =
\Phi_q[\dot z] + Y(z_\ep)\dot\sigma~,
$$
where $\Phi_q$ and $\Phi_\sigma$ denote the partial derivatives
of $\Phi$.
 So, we have
$$ 
\langle \dot z_\ep,\dot z_\ep\rangle = 
\langle Y,Y\rangle \dot \sigma^2 + 
2\dot\sigma\langle Y(z_\ep),\Phi_q[\dot z]\rangle
+ \langle \Phi_q[\dot z],\Phi_q[\dot z]\rangle
 = -\ep^2~.
\eqno(5.2)
$$
Formula (5.2) contains 
a quadratic equation on $\dot \sigma$; observe that, by
the wrong way Schwartz inequality,
the discriminant $\Delta$ of the equation (5.1) is positive:
$$
{\Delta \over 4} = 
\langle Y(z_\ep),\Phi_q[\dot z]\rangle^2
-\langle Y(z_\ep),Y(z_\ep)\rangle
\langle \Phi_q[\dot z],\Phi_q[\dot z]\rangle
+\ep^2\geq\ep^2 > 0.
\eqno(5.3)
$$
Take the solution $\sigma$ of (5.2) given by:
$$
\dot \sigma =
-\langle Y(z_\ep),Y(z_\ep)\rangle^{-1}
\left(
\langle Y(z_\ep),\Phi_q[\dot z]\rangle + {1\over 2}\sqrt{\Delta}
\right)~,
$$
where $\Delta$ is given by (5.3).
Notice that, with this choice
$$
\langle\dot z_\ep,Y(z_\varepsilon)\rangle =
\dot\sigma
\langle Y(z_\ep),Y(z_\ep)\rangle
+
\langle Y(z_\ep),\Phi_q[\dot z]\rangle=-{1\over2}\,\sqrt{\Delta} < 0~,
$$
and (5.1) is satisfied.
Observe also that the coefficients of the equation (5.2)
clearly depend continuously on $\varepsilon$.
The function $\sigma$ has to satisfy the Cauchy problem:
$$\cases{
\dot \sigma = 
-\langle Y(z_\ep),Y(z_\ep)\rangle^{-1}
\left(
\langle Y(z_\ep),\Phi_q[\dot z]\rangle
+{1\over 2}\sqrt{\Delta}~,
\right)\cr
\sigma(0) = 0~.\cr}
\eqno(5.4)
$$
Observe that, for $\ep=0$, (5.4) has the null solution, which is defined
on the whole real line.
Hence, for $\ep$ small enough,
(5.4) admits a unique solution defined on all the interval $[0,1]$.
Moreover, if $\ep$ is chosen small enough, we can also assume that
the solution $\sigma$ of (5.4) takes values in $[-\delta,\delta]$, so that
the curve $z_\ep=\Phi(\sigma,z)$ is well defined.

The construction of the map $\psi_\ep$ is done in a similar fashion,
considering
the flow $\Psi(s,q)$ of the vector field $Y$, and setting:
$$
\psi_\ep (z)(s)=z^\varepsilon(s)=\Psi(\sigma(s),z(s))~,
$$
where
$\sigma=\sigma_{z,\ep}$ is to be determined with the conditions:
$$
\sigma(0)=0,\quad
\langle \dot z^\ep,\dot
z^\ep\rangle = 0,\quad \hbox{ {\rm and} } \quad
\langle \dot z^\ep,Y(z^\ep)\rangle \leq  0~. 
$$
An argument similar to the previous case shows the existence and the
continuity 
properties of the map $\sigma$, which proves the first part of the
Proposition.

Elementary comparison theorems for ordinary differential equations 
allow to show that, for all
$z\in {\cal L}_{p,\gamma,\ep}^+$, 
the Cauchy problem (5.4) has solution defined
on the whole interval $[0,1]$. Therefore, the map $\psi_\ep$ is defined on the
whole space ${\cal L}_{p,\gamma,\ep}^+$.

Part (2) and (3) follows immediately from the construction of
$\phi_\ep$ and $\psi_\ep$.

Parts (4), and (5) follows from simple comparison theorems in O.D.E.\ 
applied to (5.4), while  part (6) follows from the
Gronwall's Lemma.
\cvd

\2

\noindent
We need also the following proposition

\2

\noindent
{\bf Proposition 5.2.}~
{\it Let $z$ be a geodesic in ${\cal L}^+_{p,\gamma}$, with $z(1)$
nonconjugate to $p$ along $z$. 
Then there exists $\ep_0 > 0$ such that for any $\ep \in ]0,\ep_0]$ there
exists one and only one 
geodesic $z_\ep \in {\cal L}^+_{p,\gamma,\ep}$, such that}
$$
\lim_{\ep\to 0}z_\ep = z_0~,
\hbox{ in the $H^{1,2}$--norm }~.
$$

\2

\noindent
{\bf Remark 5.3.}~
Notice that, if  $z_\ep$ converges to $z_0$ in the $H^{1,1}$-norm, 
using the Cauchy problem
related to the geodesic equation we immediately get that the convergence 
is also with respect to the $C^2$-norm, 
i.e., uniform up to the second derivative.

\1

\noindent
{\bf Proof.}~
Since $z(1)$ is non conjugate to $p$, the map 
$$
v \freccia \exp_p v
$$
is a local diffeomorphism between a neighborhood of $\dot z(0)$ in $T_p\M$
and a neighborhood of $z(1) = {\rm exp}_p(\dot z(0)) \in \gamma({I\!\!R})$ 
in $\M$. 
Then there exists a $C^1$--map 
$\varphi:]-\delta_0,\delta_0[ \freccia T_p\M$ such that
$$\cases{
\varphi(0) = 0,\cr
\exp_p(\dot z(0) + \varphi(\delta)) = 
\gamma(\tau(z)+\delta)~.
\cr}
\eqno(5.5)
$$
Differentiating with respect to $\delta$ and setting $\delta = 0$, we
obtain: 
$$
{\rm d}\exp_p (\dot z(0))[\varphi'(0)] = 
\dot\gamma(\tau(z))~.
\eqno(5.6)
$$
The following lemma is needed.

\2

\noindent
{\bf Lemma 5.4.}~
{\it Fix $v_0 \in T_p\M$ lightlike and future pointing. 
Set $V = {\rm d}\exp_p(v_0)[v]$. 
Assume that $V$ is timelike and future pointing. Then
$\langle v,v_0\rangle < 0$.}

\1

\noindent
{\bf Proof.}~
Denote by $z$ the geodesic such that $z(0) = p$ and $\dot z(0) = v_0$. 
As known, since $v_0 = \dot z(0)$,  
${\rm d}\exp_p(v_0)[v]$ is given by $Z(1)$, where $Z$ 
 is the unique Jacobi field along $z$ such that $Z(0) = 0$ and $D_s Z(0) = v$.
Since $\zeta(s)  = s\dot z(s)$ is the unique Jacobi field along $z$ such that 
$\zeta(0) = 0$ and $D_s \zeta (0) = \dot z(0) = v_0$, we have
$$
{\rm d}\exp_p(v_0)[v_0] = \dot z(1)~.
\eqno(5.7)
$$
Now, by the Gauss Lemma (cf. [BEE]), for any $v \in T_p\M$, we have:
$$
\langle 
{\rm d}\exp_p(v_0)[v_0],
{\rm d}\exp_p(v_0)[v]\rangle = 
\langle v _0,v\rangle~.
\eqno(5.8)
$$
By (5.7), $V_0 = {\rm d}\exp_p(v_0)[v_0]$ is lightlike and future
pointing. Indeed, $\dot z(1)$ is lightlike and future pointing, since 
$\dot z(0) = v_0$ is lightlike and future pointing. 
Moreover, $V = {\rm d}\exp_p(v_0)[v]$ is timelike and future pointing
by assumption, therefore 
$\langle V_0,V\rangle < 0$.
Then, by (5.8) the proof is complete.
\cvd

\2

\noindent
Now, let us go back to the proof of Proposition 5.3. 

\2

Since $\dot\gamma(\tau(z(1)))$ is timelike and future pointing, by (5.6)
and Lemma 5.4 we get
$$
\langle \varphi'(0),\dot z(0)\rangle < 0~.
\eqno(5.9)
$$
By (5.9), since $\varphi(0) = 0$ 
and $\langle\dot z(0),\dot z(0)\rangle = 0$, up to the choice of a smaller
$\delta_0$, we immediately obtain
$$
\langle 
\dot z(0) + \varphi(\delta),\dot z(0) + \varphi(\delta)\rangle < 0~,
\quad 
\forall \delta > 0~.
\eqno(5.10)
$$
Moreover, since $\varphi(0) = 0$, for any $\delta$ sufficiently small, 
$$
\langle \dot z(0) + \varphi(\delta),Y(z(0))\rangle < 0~.
\eqno(5.11)
$$
Then we can conclude the proof taking $\ep_0 = \ep_0(\delta_0)$ and 
$$
\ep = \ep(\delta) = 
\sqrt{
-\langle 
\dot z(0) + \varphi(\delta),\dot z(0) + \varphi(\delta)
\rangle
}~,
\eqno(5.12)
$$
which is well defined because of (5.10).

Indeed the geodesic $z_\ep$ such that $z_\ep(0) = p$ and 
$\dot z_\ep (0) = z(0) + \varphi(\delta)$ 
is in ${\cal L}^+_{p,\gamma,\ep}$, since, 
by (5.12), it is $\langle\dot z_\ep,\dot z_\ep\rangle = -\ep^2$. 

Moreover, by (5.11) $\dot z_\ep (0)$ is future pointing, so
that $z_\ep(s)$ is timelike and future pointing for any $s$.
Finally, the $C^2$ convergence of $z_\ep$ to $z$ is obvious by 
the continuous
dependence of the solutions of differential equations on its
data.
\cvd
\2

\noindent 
We conclude the section with an useful  result  for the proof of 
Theorems~1.4 and 1.6.

\2

\noindent
{\bf Proposition 5.5.}~
{\it Suppose that $\tau$ is pseudo-coercive on ${\cal L}_{p,\gamma}^+$.
Let $(\ep_n)_{n\in\N}$ be a sequence of positive numbers converging to $0$ and
$z_n\in {\cal L}_{p,\gamma,\ep_n}^+$ be  a sequence of curves such that:}
$$\sup_n\tau(z_n)=\overline c<+\infty. $$
{\it 
Then, denoting by $l(z_n)$ the length of $z_n$ with respect to the Riemannian
metric {\rm (2.4)}, it is:}
$$\sup_n l(z_n)<+\infty. $$
{\it 
Moreover there exists $K$, compact subset of $\M$, such that 
$z_n([0,1]) \subset K$ for any $n$.}
\medskip
\noindent{\bf Proof.} \enspace
Denote by $\tilde z_n$ the sequence:
$$
\tilde z_n=\psi_{\ep_n}(z_n)\in \hat{\cal L}_{p,\gamma}^+.
\eqno(5.13)
$$
By (5) of Proposition 5.1 it is 
$\tau(\tilde z_n)\le\tau(z_n)\le \overline c$. 
Then, by the same arguments
used in the proof of Lemma 2.6, 
we see that the pseudo-coercivity of $\tau$ implies that 
$$
l(\tilde z_n)\le \tilde c<+\infty \hbox{ for any $n$,}
\eqno{(5.14)}
$$
and  there exists a compact subset of $\M$ 
containing the images of all the $\tilde z_n$'s.
Moreover, since $z_n=\phi_{\ep_n}(\psi_{\ep_n}(z_n))$, 
by (6) of Proposition 5.1 
there exists $M > 0$ such that 
$$d_2(\tilde z_n,z_n)\le M\cdot\ep_n, $$
Therefore it follows that $l(z_n)$ is bounded 
and there exists a compact subset of 
$\M$ containing the images of the $z_n$'s.  \cvd 

\vfil
\eject

\noindent
{\bf 6. The limit process and the Morse Relations on 
${\cal L}^+_{p,\gamma}.$}

\noindent
In this section we shall prove Theorems 1.6 and 1.4.
\1
\noindent
{\bf Proof of Theorem 1.6.}~
 Let $z_n$ be as
in the statement of Theorem~1.6.
Since $\tau(z_n)\le c$ for all $n$, by Proposition~5.5 there exists
a compact subset $K$ of $\M$ and a positive constant $C$ such that:
$$
z_n([0,1])\subset K,\quad\hbox{\rm and}\quad l(z_n)\le C,
\qquad\forall\,n\in\N.
$$
Then, the proof is obtained passing to the limit as $\varepsilon\to0$
in the Cauchy problem related to the geodesic equation satisfied
by the $z_n$'s.
\cvd
\1

\noindent
{\bf Proof of Theorem 1.4.}~
Let $c$ be a regular value for $\tau$ on ${\cal L}^+_{p,\gamma}$, i.e., 
$\tau^{-1}(c)\cap {\cal L}^+_{p,\gamma}$ does 
not contain geodesics. 
By assumption 2), all the geodesics in  
$\tau^{-1}(c) \cap {\cal L}^+_{p,\gamma}$ 
are isolated. 
A simple compactness argument shows that they are finite. 
By Proposition 5.2 there exists a positive number $\ep(c)$ such that 
for any geodesic $z_i$ in 
$\tau^{-1}(c)\cap {\cal L}^+_{p,\gamma}$ 
and for any $\ep \in ]0,\ep(c)]$, 
there exists an unique geodesic
$z^i_\ep \in \tau^{-1}(c) \cap {\cal L}^+_{p,\gamma,\ep}$ approaching 
$z_i$ for any $i = 1,\dots,k$.
Choose $\ep (c)\leq \ep_0(c)$ given by Proposition 5.1 and denote 
(for any $\ep \in\,]0,\ep_{0}]$) by $c_\ep$
the minimal real number such that
$$
\phi_\ep(\tau^c\cap {\cal L}^+_{p,\gamma}) 
\subset \tau^{c_\ep}\cap {\cal L}^+_{p,\gamma,\ep}~,
$$
where $\phi_\ep$ is defined in Proposition 5.1. 

If $\ep (c)$ is sufficiently small, any $c_\ep$ is a regular value for
$\tau$ on ${\cal L}^+_{p,\gamma,\ep}$ 
for all $\ep\in\,[0,\ep(c)]$ and for any
geodesic in ${\cal L}^+_{p,\gamma,\ep}$ 
"correspond" to a unique geodesic on
${\cal L}^+_{p,\gamma}$ (having the same geometric index) 
(cf. Proposition 5.2 and Theorem 1.7).

Moreover, choosing $\ep(c)$ small enough, 
by the pseudo-coercivity of $\tau$  on
${\cal L}_{p,\gamma}^+$ we have the existence of a compact subset $K=K(c)$
of $\M$ and of a positive constant $L=L(c)$ such that $z([0,1])\subset K$
and
$l(z)\le L(c)$ for all $\ep \in [0,\ep(c)$] and for all
$z\in\tau^{c_\ep}\cap {\cal L}_{p,\gamma,\ep}^+$ (cf. Proposition 5.5).

This allow us to use the 
curve shortening method at every level 
$b \leq c_{\ep}$ and Propositions 3.3 and 3.4.

Arguing as in the proof of Theorem~4.4 and using Theorem~4.1, we can write
the following Morse relations, valid for every $\ep \in ]0,\ep(c)]$ 
and every coefficients field ${\cal K}$:
$$
\sum_{z_\ep \in {\cal G}_{p,\gamma,\ep}^{+,c_\ep}}
\lambda^{\mu (z_\ep)} = 
{\cal P}_\lambda(\tau^{c_\ep}\cap {\cal L}_{p,\gamma,\ep}^+,{\cal K}) + 
(1+\lambda){\cal S}_\ep(\lambda)~,
\eqno(6.1)
$$
where 
${\cal G}_{p,\gamma,\ep}^{+,d} = {\cal G}_{p,\gamma}^+\cap
\tau^{d}$.

Now choose a monotone sequence $c_m$ of regular values for $\tau$ on 
${\cal L}_{p,\gamma}^+$ such that $c_m \to +\infty$.
For any $m$ let $\ep_m = \ep(c_m)$ as above.
Let $d_m$ be the minimal real number such that
$$\phi_\ep(\tau^{c_m} \cap {\cal L}_{p,\gamma}^+) \subset \tau^{d_m} \cap
{\cal L}_{p,\gamma,\ep}^+ \hbox{  for any  } \ep \in [0,\ep_m].
$$
\noindent 
(Note that $d_m \geq c_m$). By (6.1) and Proposition 5.1 we deduce
$$
\sum_{z\in{\cal G}^{+,d_{m}}_{p,\gamma}}\lambda^{\mu(z)}
= 
{\cal P}_\lambda(\psi_{\ep_m}(\tau^{d_m} \cap 
{\cal L}_{p,\gamma,\ep_m}^+),{\cal K}) +
(1+\lambda){\cal S}^{'}_{m}(\lambda)~,
$$
where ${\cal S}^{'}_{m}$ is a polynomial 
with non negative integer coefficients.

By the exactness in singular homology of the pair 
${\cal L}^+_{p,\gamma},\psi_{\ep_m}(\tau^{d_m} \cap 
{\cal L}_{p,\gamma,\ep_m}^+))$, there exists a formal series $R_m$
(with coefficients in $\N \cup \{+\infty\}$) such that (cf. e.g. [MW])
$$
{\cal P}_\lambda
(\psi_{\ep_m}(\tau^{d_m} \cap {\cal L}_{p,\gamma,\ep_m}^+))
+
{\cal P}_\lambda
({\cal L}^+_{p,\gamma}
,\psi_{\ep_m}(\tau^{d_m} \cap {\cal L}_{p,\gamma,\ep_m}^+)
=
{\cal P}_\lambda({\cal L}_{p,\gamma}^+) +
(1+\lambda)R_m(\lambda)~. 
$$
Then, there exists a formal
series $S_m$ such that
$$
\sum_{z\in{\cal G}^{+,d_m}_{p,\gamma}}\lambda^{\mu(z)}
+ {\cal P}_\lambda({\cal L}_{p,\gamma}^+,
\psi_{\ep_m}(\tau^{d_m} \cap {\cal L}_{p,\gamma,\ep_m}^+))=
{\cal P}_\lambda({\cal L}_{p,\gamma}^+) + 
(1+\lambda){\cal S}_m(\lambda)~.
\eqno(6.2)
$$

Let $N(l,m)$ be the number of lightlike geodesics in 
$\psi_{\ep_m}(\tau^{d_m} \cap {\cal L}_{p,\gamma,\ep_m}^+)$ having 
geometric index equal to $l$. By Proposition 5.1, the subsets 
$\psi_{\ep_m}(\tau^{d_m} \cap {\cal L}_{p,\gamma,\ep_m}^+)$ are
ordered by inclusion. Then $N(l,m)$ is nondecreasing in $m$ and 
tends, as $m\to +\infty$, to the number $N(l)$ of the
lightlike geodesics in 
${\cal L}_{p,\gamma}^+$ having geometric index equal to $l$. 
Since ${\N}\cup \{+\infty\}$ is compact (with respect to its usual
convergence), a diagonalization argument shows the existence of a
subsequence $(m_k)_{k\in{\N}}$ such that, for any $l \in{\N}$ the
sequences $(b_{l,m_k})$ of the formal series $S_{m_k}$ in (6.2)
converges to $b_l \in {\N}\cup\{+\infty\}$. 
Then, up to considering subsequences, every coefficient $b_{l,m}$ of $S_m$
is convergent to $b_l$.
We shall prove (1.4) arguing for any 
coefficient $l\in{\N}$.
If $N(l) = +\infty$, either 
the $l$--th coefficient $\beta_l$ of 
${\cal P}_{\lambda}({\cal L}^+_{p,\gamma},{\cal K})$ 
is equal to $+\infty$, or 
at least one between $b_{l-1}$ and $b_l$ is equal to $+\infty$. 
In any case
$$
N(l) = \beta_l + b_{l-1} + b_l~,
\eqno(6.3)
$$
obtaining (1.4) relatively to the $l$--th coefficient.

Assume now that $N(l) < +\infty$. Let
$$
b_{*} = \max\{\tau(z):z\in {\cal G}^+_{p,\gamma}, \mu(z) = q\}~.
\eqno(6.4)
$$
By (6.2), in order to prove (6.3), it suffices to show the vanishing
of the Betti number: 
$$
\beta_l(
{\cal L}_{p,\gamma}^+,
\psi_{\ep_m}(\tau^{d_m} \cap {\cal L}_{p,\gamma,\ep_m}^+)) = 0, \quad
\hbox{ $\forall m$ such that $c_m > b_{*}$ }~.
\eqno(6.5)
$$
Assume by contradiction that (6.5) does not hold. 
Let $\Delta_m$ be a nontrivial element of the homology group
$H_l({\cal L}_{p,\gamma}^+,
\psi_{\ep_m}(\tau^{d_m} \cap {\cal L}_{p,\gamma,\ep_m}^+))$
and let $K_m$ be its compact support.
Now for any $\ep \in ]0,\ep_m]$, by Proposition 5.1,
there exists $\mu_m > 0$ (infinitesimal as $\ep_m$ tends to $0$), 
such that $$\eqalign{
\psi_{\ep_m}(\tau^{d_m} \cap {\cal L}_{p,\gamma,\ep_m}^+) &\subset
\psi_{\ep}(\tau^{d_m} \cap {\cal L}_{p,\gamma,\ep}^+) \subset \cr
&\subset\psi_{\ep_m}(\tau^{d_m+\mu_m} 
\cap {\cal L}_{p,\gamma,\ep_m}^+) \subset
\psi_{\ep}(\tau^{d_m+\mu_m} \cap {\cal L}_{p,\gamma,\ep}^+). \cr}
$$
Now, choosing $\ep_m$ small enough, we can assume that
there are no geodesics in the strip
$\tau^{-1}([d_m,d_m+\mu_m]) \cap {\cal L}^+_{p,\gamma,\ep}$,  
for all $\ep \in ]0,\ep_m]$.
Then, if $\ep_m$ is small, 
$\psi_{\ep_m}(\tau^{d_m} \cap {\cal L}_{p,\gamma,\ep_m}^+)$
is a strong deformation retract of  $\psi_{\ep_m}(\tau^{d_m+\mu_m} \cap
{\cal L}_{p,\gamma,\ep_m}^+)$ and
$\psi_{\ep}(\tau^{d_m} \cap {\cal L}_{p,\gamma,\ep}^+)$
is a strong deformation retract of
$\psi_{\ep}(\tau^{d_m+\mu_m} \cap {\cal L}_{p,\gamma,\ep}^+)$
for any $\ep \in ]0,\ep_m]$.
(Recall that $Y \subset X$  is a strong deformation retract 
of $X$ if there exists a continuous map
$H:[0,1]\times X$ such that $H(0,\cdot)$ is the identity on $X$, 
$H(s,\cdot)$ is the identity on $Y$ for all $s$, and
$H(1,X) \subset Y$).

Then, by standard techniques in Algebraic Topology we have that, 
for any $k \in \N$ 
$$
i_{k}^{*} : H_k({\cal L}^+_{p,\gamma},\psi_{\ep_m}(\tau^{d_m} \cap 
{\cal L}_{p,\gamma,\ep_m}^+))
\longrightarrow  
H_k({\cal L}^+_{p,\gamma},\psi_{\ep}(\tau^{d_m} \cap 
{\cal L}_{p,\gamma,\ep}^+))
$$
(where $i$ denotes the inclusion map) is an isomorphism. Therefore, 
there exists 
$\Delta_\ep \in H_l({\cal L}^+_{p,\gamma,\ep},\psi_{\ep}(\tau^{d_m}
\cap {\cal L}_{p,\gamma}^+)) \setminus \{0\}$ with support $K_m$. 
Finally, choose
$$
C_m > \sup\{\tau(z):z \in K_m \cap {\cal L}^+_{p,\gamma}\}, \quad\quad
\hbox{ $C_m$ regular value for $\tau$ on ${\cal L}_{p,\gamma}^+$.}
$$
(Clearly, $C_m$ can be chosen larger than $d_m$). 
Using the exactness of the triple 
$({\cal L}^+_{p,\gamma},\psi_{\ep}(\tau^{C_m} 
\cap {\cal L}_{p,\gamma,\ep}^+),
\psi_{\ep}(\tau^{d_m} \cap {\cal L}_{p,\gamma,\ep}^+)),$
gives the existence of 
$$\Gamma_\ep \in
H_l(\psi_{\ep}(\tau^{C_m} \cap {\cal L}_{p,\gamma,\ep}^+),
\psi_{\ep}(\tau^{d_m} \cap {\cal L}_{p,\gamma,\ep}^+)) \setminus \{0\}$$
with support $K_m$.

Since $\psi_{\ep}$ is an homeomorphism there exists $$\hat \Gamma_\ep \in
H_l(\tau^{C_m} \cap {\cal L}_{p,\gamma,\ep}^{+},
\tau^{d_m} \cap {\cal L}_{p,\gamma,\ep}^+) \setminus \{0\}$$
with support $K_m$.
Using the curve shortening method and the classical Morse Theory nearby
critical points shows 
the existence of  $\hat{\ep} \in ]0,\ep_m]$ such that,
for any $\ep \in \hat{\ep}$ we have the existence of a geodesic
$z_\ep \in \tau^{-1}([d_m,C_m]) \cap {\cal L}_{p,\gamma,\ep}^+)$ 
having index $l$ (see Theorem 4.4).
Finally sending $\ep$ to $0$, by Theorems 1.6 and 1.7
we obtain the existence of a geodesic 
$z$ in ${\cal L}_{p,\gamma}^+$ such that 
$$
\mu(z) = q,\quad \tau(z) \in [d_m,C_m]~.
$$
In particular $\tau(z) \geq d_m \geq c_m > b_{*}$,
in contradiction with (6.4).
\cvd

\vfil
\eject

\noindent
{{\bf References}}

\1

\item{[BEE]}
BEEM J.K., EHRLICH P.E., EASLEY K.L.,
{\it Global Lorentzian Geometry}, 
Marcel Dekker: New York, 1996.

\item{[Br]}
BREZIS H., 
{\it Analyse fonctionelle}, 
Masson: Paris, 1984.

\item{[C]}
CHANG K.\ C.,
{\it Infinite Dimensional Morse Theory and Multiple Solutions Problems},
Birkhauser: Boston, 1993.

\item{[Ek]}
EKELAND I., 
On the variational principle, 
{\it J. Math. Anal. Appl.} {\bf 47}, 324--353 (1974).

\item{[GM]}
GIANNONI F., MASIELLO A., 
On a Fermat principle in General Relativity. 
A Ljusternik--Schnirelmann theory for light rays, 
{\it Ann. Mat. Pura Appl. (IV)}, {\bf CLXXIV} 161--207 (1998).

\item{[GMP1]}
GIANNONI F.,  MASIELLO A., PICCIONE P., 
A variational theory for light rays in stably causal Lorentzian
manifolds: regularity and multiplicity results, 
{\it Comm.\ Math.\ Phys.} {\bf 187}, 375--415 (1997).

\item{[GMP2]}
GIANNONI F.,  MASIELLO A., PICCIONE P., 
A Morse Theory for light rays on stably causal Lorentzian manifolds, 
{\it Ann.\ Inst.\ H.\ Poincar\'e, Physique Theorique} {\bf 69}, 359--412
(1998). 

\item{[GMP3]}
GIANNONI F.,  MASIELLO A., PICCIONE P., 
A timelike extension of Fermat's principle in General Relativity and
applications, 
{\it Calc.\ Var.}, {\bf 6}, 263-283 (1998).

\item{H}
HELFER A., 
Conjugate points on spacelike geodesics or pseudo--selfadjoint
Morse--Sturm--Liouville operators, 
{\it Pac. J. Math}, {\bf 164} 321--340 (1994).

\item{[K]}
KOVNER I.,
Fermat principles in arbitrary gravitational fields, 
{\it Astrophys.\ J.} {\bf 351}, 114--120 (1990).

\item{[La]} 
LANG, S.,  Differential Manifolds, Springer-Verlag,
Berlin, 1985.

\item{[L]}
LOMBARDI M.,
An application of the topological degree to gravitational lenses, 
{\it Modern Phys.\ Letters A}, {\bf 13} 83-86 (1997) 

\item{[Ma]}
MASIELLO A.,
{\it Variational methods in Lorentzian Geometry}.
Pitman Research Notes in Mathematics Series {\bf 309}. Longman Ed.: 
London, 1994.

\item{[MW]}
MAWHIN J., WILLEM M,
{\it Critical point theory and Hamiltonian systems},
Springer Verlag: New York--Berlin 1988. 

\item{[MPT]}
MERCURI F., PICCIONE P., TAUSK D.V., 
Stability of the focal and geometric index in semi-Riemannian geometry via
the Maslov index, 
Technical Report RT-MAT ../99, Dep.\ Mathematics, Univ. S\~ao Paulo, Brazil.

\item{[Mi]}
MILNOR J.
{\it Morse Theory}. 
Ann.\ Math.\ Stud.\ {\bf 51}, Princeton University Press: Princeton, 1963.

\item{[N]}
NASH J.,
The embedding problem for Riemannian manifolds, 
{\it Ann.\ Math.} {\bf 63}, 20--63 (1956).

\item{[ON]}
O'NEILL B., 
{\it Semi--Riemannian Geometry with applications to Relativity}. 
Ac.\ Press: New York--London, 1983.

\item{[Pa1]}
PALAIS R.S., {\it Foundations of Global Nonlinear Analysis},
W.\ A.\ Benjamin, 1968.

\item{[Pa2]}
PALAIS R.S., Morse Theory on Hilbert manifolds, 
{\it Topology} {\bf 2}, 299--340 (1963).

\item{[Pe]}
PERLICK V.,
On Fermat's principle in General Relativity: I. The general case,
{\it Class.\ Quantum Grav.} {\bf 7}, 1319--1331 (1990).

\item{[Pt1]}
PETTERS A., 
Multiplane gravitational lensing I. Morse Theory and image counting,
{\it J.\ Math.\ Phys.} {\bf 36}, 4263--4275 (1995).

\item{[Pt2]}
PETTERS A., 
Multiplane gravitational lensing II. Global Geometry of caustics,
{\it J.\ Math.\ Phys.} {\bf 36}, 4276--4295 (1995).

\item{[Sc]}
SCHNEIDER P.,
A new formulation of gravitational lens theory, time--delay and Fermat's
principle, 
{\it Astr.\ Astrophys.} {\bf 143}, 413--420 (1985).

\item{[SEF]}
SCHNEIDER P., EHLERS J., FALCO E.E.
{\it Gravitational Lenses}, 
Springer-Verlag: New York, 1992.

\item{[Sp]}  SPANIER H., 
{\it Algebraic Topology},
Mc Graw Hill. New York, 1966.

\item{[St]}
STRUWE M. 
{\it Variational Methods}, 
Springer Verlag: New York--Berlin 1996.
\vfill
\noindent{\bf Note.} \enspace {\it The shortening method described in Section~3 is
illustrated in the five pictures appearing in the next pages.}
\eject
\end